\documentclass[graybox]{svmult}

\usepackage{type1cm}        % activate if the above 3 fonts are
\usepackage{makeidx}         % allows index generation
\usepackage{graphicx}        % standard LaTeX graphics tool
\usepackage{multicol}        % used for the two-column index
\usepackage[bottom]{footmisc}% places footnotes at page bottom

\usepackage{newtxtext}       % 
\usepackage{newtxmath}       % selects Times Roman as basic font

\usepackage{bbm}
\usepackage{subcaption}
\newcommand\ddfrac[2]{\frac{\displaystyle #1}{\displaystyle #2}}

\makeindex             % used for the subject index
\begin{document}

\title*{Strategic Learning for Active, Adaptive,  and Autonomous Cyber Defense} 
 %\titlerunning{Strategic Learning for Adaptive Cyber Deception}
\author{Linan Huang \and 
Quanyan Zhu}
\institute{Linan Huang\at Department of Electrical and Computer Engineering, New York University, 2 MetroTech Center, Brooklyn, NY, 11201, USA, \email{lh2328@nyu.edu}
\and Quanyan Zhu \at Department of Electrical and Computer Engineering, New York University, 2 MetroTech Center, Brooklyn, NY, 11201, USA, \email{qz494@nyu.edu}
}
\maketitle

\abstract{
%Each chapter should be preceded by an abstract (no more than 200 words) that summarizes the content. 
The increasing instances of advanced attacks call for a new defense paradigm that is active, autonomous, and adaptive, named as the \texttt{`3A'} defense paradigm.
This chapter introduces three defense schemes that actively interact with attackers to increase the attack cost and gather threat information, i.e., defensive deception for detection and counter-deception, feedback-driven Moving Target Defense (MTD), and adaptive honeypot engagement. 
Due to the cyber deception, external noise, and the absent knowledge of the other players' behaviors and goals, these schemes possess three progressive levels of information restrictions, i.e., from the parameter uncertainty, the payoff uncertainty, to the environmental uncertainty.  
To estimate the unknown and reduce the uncertainty, we adopt three different strategic learning schemes that fit the associated information restrictions. 
All three learning schemes share the same feedback structure of sensation, estimation, and actions so that the most rewarding policies get reinforced and converge to the optimal ones in autonomous and adaptive fashions. 
This work aims to shed lights on proactive defense strategies, lay a solid foundation for strategic learning under incomplete information, and quantify the tradeoff between the security and costs. 
%193 words. 
}

\section{Introduction}
Recent instances of \texttt{WannaCry} ransomware, \texttt{Petya} cyberattack, and \texttt{Stuxnet} malware have demonstrated the trends of modern attacks and the corresponding new security challenges as follows. 
\begin{itemize}
\item \textbf{Advanced}: Attackers  leverage sophisticated attack tools to invalidate the off-the-shelf defense schemes such as the firewall and intrusion detection systems. 
\item \textbf{Targeted}:  Unlike automated probes, targeted attacks conduct thorough research in advance to expose the system architecture, valuable assets, and  defense schemes. 
\item \textbf{Persistent}: Attackers can restrain the adversary's behaviors and bide their times to launch critical attacks. They are persistent in achieving the goal. 
\item \textbf{Adaptive}: Attackers can learn the defense strategies and unpatched vulnerabilities during the interaction with the defender and tailor their strategies accordingly. 
\item \textbf{Stealthy and Deceptive}: Attackers conceal their true intentions and disguise their claws to evade detection. The adversarial cyber deception endows attackers an information advantage over the defender. 
\end{itemize}

Thus, defenders are urged to adopt active, adaptive, and autonomous defense paradigms to deal with the above challenges and proactively protect the system prior to the attack damages rather than passively compensate for the loss.  
In analogy to the classical \textbf{Kerckhoffs's principle} in the 19th century that attackers know the system, we suggest a new security principle for modern cyber systems as follows:  
\begin{svgraybox}
\textbf{Principle of 3A Defense}: 
A cyber defense paradigm is considered to be insufficiently secure if its effectiveness
relies on 
\begin{itemize}
\item Rule-abiding human behaviors. 
\item A perfect protection against vulnerabilities and  a perfect prevention from system penetration.
\item A perfect knowledge of attacks. 
\end{itemize}
\end{svgraybox}
Firstly, $30\%$ of data breaches are caused by privilege misuse and error by insiders according to Verizon's data breach report in $2019$ \cite{Jeff2000}. Security administration does not work well without the support of technology, and autonomous defense strategies are required to deal with the increasing volume of sophisticated attacks. 
Secondly, systems always have undiscovered vulnerabilities or unpatched vulnerabilities due to the long supply chain of uncontrollable equipment providers \cite{shackleford2015combatting} and the increasing complexities in the system structure and functionality. 
Thus, an effective paradigm should assume a successful infiltration and pursue strategic securities through interacting with intelligent attackers. 
Finally, due to adversarial deception techniques and external noises, the defender 
cannot expect a perfect attack model with predicable behaviors. The defense mechanism should be robust under incomplete information and adaptive to the evolution of attacks. 

\begin{figure*}[h]
\sidecaption[t]
\centering
\includegraphics[width=0.63 \textwidth]{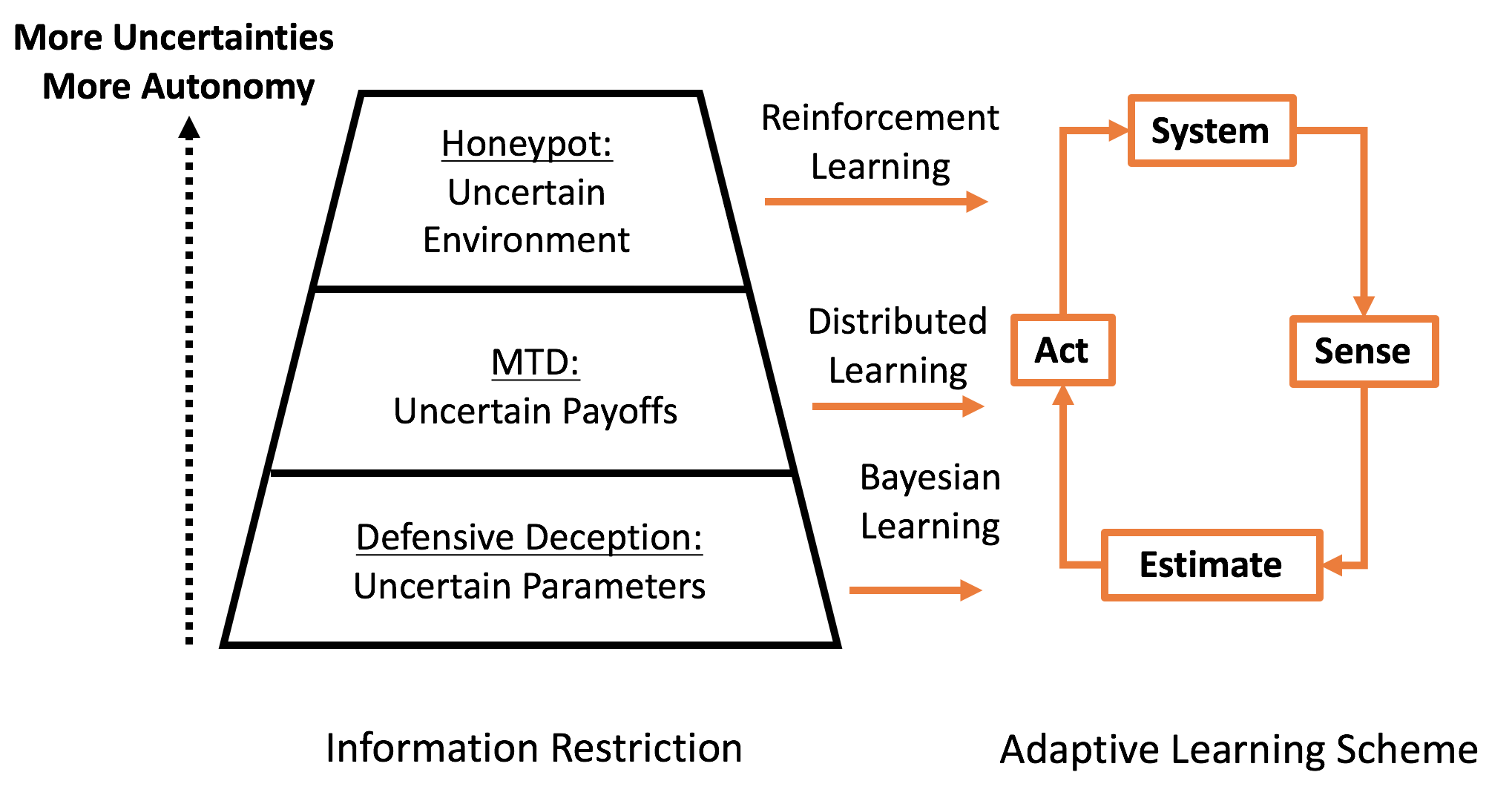}
\caption{
The left part of the figure describes the degree of information restriction. From bottom to top, the defense scheme becomes more autonomous and relies less on an exact attack model, which also result in more uncertainties. 
The right part is the associated feedback learning schemes. 
}
\label{fig: wholepic}
\end{figure*}

In this chapter, we illustrate three active defense schemes in our previous works, which are designed based on the new cyber security principle. 
They are defensive deception for detection and counter-deception \cite{huang2019adaptive,huang2018analysis,APTjournal} in Section \ref{sec:Type}, feedback-driven Moving Target Defense (MTD) \cite{zhu2013game} in Section \ref{sec:MTD}, and adaptive honeypot engagement \cite{huangHoneypot} in Section \ref{sec:honeypot}. 
All three schemes is of incomplete information, and we arrange them based on three progressive levels of information restrictions as shown in the left part of Fig. \ref{fig: wholepic}. 

 The first scheme in Section \ref{sec:Type} considers the obfuscation of characteristics of known attacks and systems through a random parameter called the player's \textit{type}. The only uncertainty origins from the player's \textit{type}, and the mapping from the type to the utility is known deterministically. 
 The MTD scheme in Section \ref{sec:MTD} considers unknown attacks and systems whose utilities are completely uncertain, while the honeypot engagement scheme in Section \ref{sec:honeypot} further investigates environmental uncertainties such as the transition probability, the sojourn distribution, and the investigation reward.

To deal with these uncertainties caused by different information structures, we suggest three associated learning schemes as shown in the right part of Fig. \ref{fig: wholepic}, i.e., Bayesian learning for the parameter estimation, distributed learning for the utility acquisition without information sharing, and reinforcement learning for the optimal policy obtainment under the unknown environment. 
 All three learning methods form a feedback loop that strategically incorporates the samples generated during the interaction between attackers and defenders to persistently update the beliefs of known and then take actions according to current optimal decision strategies. 
The feedback structure makes the learning adaptive to behavioral and environmental changes. 

Another common point of these three schemes is the quantification of the tradeoff between security and the different types of cost.
In particular, the costs result from the attacker's identification of the defensive deception, the system usability, and the risk of attackers penetrating production systems from the honeynet, respectively. 

\subsection{Literature}
The idea of using deceptions defensively to detect and deter attacks has been studied theoretically as listed in the taxonomic survey \cite{pawlick2017game},  implemented to 
 the Adversarial Tactics, Techniques and Common Knowledge
(ATT\&CK$ ^{TM}$) adversary model system  \cite{stech2016integrating}, and tested in the real-time cyber-wargame experiment \cite{heckman2013active}. 
Many previous works imply the similar idea of type obfuscation, e.g., creating social network avatars (fake personas) on the major social networks 
\cite{gomez2018r}, implementing honey files for ransomware actions \cite{virvilis2014changing}, and disguising a production system as a honeypot to scare attackers away \cite{pawlick2018modeling}. 

Moving target defense (MTD) allows dynamic security strategies to limit the exposure of vulnerabilities and the effectiveness of the attacker's reconnaissance by increasing complexities and costs of attacks \cite{jajodia2011moving}. 
To achieve an effective MTD, \cite{kc2003countering} proposes the instruction set and the address space layout randomization, \cite{clark2012deceptive} studies the deceptive routing against jamming in multi-hop relay networks, and \cite{maleki2016markov} uses the Markov chain to model the MTD process and discusses the optimal strategy to balance the defensive benefit and the network service quality. 

The previous two methods use the defensive deception to protect the system and assets. 
To further gather threat information, the defender can implement honeypots to lure attackers to conduct adversarial behaviors and reveal their TTPs in a controlled and monitored environment. 
Previous works \cite{hecker2012methodology, la2016deceptive} have investigated the adaptive honeypot deployment to effectively engage attackers without their  notices. 
The authors in recent work \cite{PawlickNZ17} proposes a continuous-state Markov Decision Process (MDP) model and focuses on the optimal timing of the attacker ejection. 

%\subsubsection{Game of Incomplete Information and Strategic Learning} 
Game-theoretic models are natural frameworks to capture the multistage interaction between attackers and defenders. Recently, game theory has been applied to  different sets of
security problems, e.g., Stackelberg and signaling games for
deception and proactive defenses
\cite{pawlick_stackelberg_2016,zhu2013game,zhu2013deployment,zhu2013hybrid,zhu2012interference,clark2012deceptive,zhu2012game,zhu2012deceptive,zhu2010stochastic}, network games for cyber-physical security
\cite{xu2017secure,xu_game-theoretic_2017,xu_cross-layer_2016,farooq2019modeling,xu2015cyber,huang2017large,chen2017dynamic,miao2018hybrid,yuan2013resilient,Rass&Zhu2016},
 dynamic games for adaptive defense
\cite{zhu2010dynamic,zhang2017strategic,huang2018gamesec,huang2018PER,huang2019adaptive,pawlick2015flip,farhang2014dynamic,zhu2009dynamic,zhu2010network,zhu2010heterogeneous}, and mechanism design theory for security
\cite{chen_security_2017,zhang_bi-level_2017,zhang_attack-aware_2016,casey2015compliance,hayel2015attack,hayel2017epidemic,zhu2012guidex,zhu2012tragedy,zhu2009game}.

%\cite{huang2017large,zhu2018multi,huangcontrol}. 
%The FlipIt game \cite{van2013flipit}, the signaling game \cite{zhang2017strategic}, and the cascading game-of-games \cite{pawlick2018istrict}, further incorporate incomplete information structure caused by cyber deception. 

Information asymmetry among the players in network security is a challenge to deal with. The information asymmetry can be either leveraged or created by the attacker or the defender for achieving a successful cyber deception. For example, techniques such as honeynets \cite{carroll2011game,zhu2013deployment}, moving target
defense \cite{zhu2013game,jajodia2011moving,huang2019adaptive}, obfuscation \cite{pawlick2016stackelberg,zhang_dynamic_2017,farhang2015phy,zhang2018distributed},
and mix networks \cite{zhang2010gpath} have been introduced to create difficulties for attackers to map out the system information.

To overcome the created or inherent uncertainties of networks, many works have studied the strategic learning in security games, e.g., Bayesian learning for unknown adversarial strategies \cite{garnaev2015security}, heterogeneous and hybrid distributed learning \cite{zhu2010heterogeneous,zhu2011distributed}, multiagent reinforcement learning for intrusion detection \cite{servin2008multi}. 
Moreover, these learning schemes are combined to achieve better properties, e.g., distributed Bayesian learning \cite{djuric2012distributed}, Bayesian reinforcement learning \cite{chalkiadakis2003coordination}, and distributed reinforcement learning \cite{chen2015distributed}.

\subsection{Notation}
Throughout the chapter, we use calligraphic letter $\mathcal{A}$ to define a set and $|\mathcal{A}|$ as the cardinality of the set. 
Let $\bigtriangleup \mathcal{A}$ represent the set of probability distributions over $\mathcal{A}$. If set $\mathcal{A}$ is discrete and finite, $\bigtriangleup \mathcal{A}:=\{p:\mathcal{A} \mapsto R_{+} | \sum_{a\in \mathcal{A}} p(a)=1\}$, otherwise, $\bigtriangleup \mathcal{A}:=\{p:\mathcal{A} \mapsto R_{+} | \int_{a\in \mathcal{A}} p(a)=1\}$. 
Row player $P_1$ is the defender (pronoun `she') and $P_2$ (pronoun `he') is the user (or the attacker) who controls the column of the game matrix. 
Both players want to maximize their own utilities. 
 The indicator function $\mathbf{1}_{\{x=y\}}$ equals one if $x = y$, and zero if $x\neq y$. 
\subsection{Organization of the Chapter}
The rest of the paper is organized as follows. In Section \ref{sec:Type}, we elaborate defensive deception as a countermeasure of the adversarial deception under a multistage setting where Bayesian learning is applied for the  parameter uncertainty. Section \ref{sec:MTD} introduces a multistage MTD framework and the uncertainties of payoffs result in distributed learning schemes. 
Section \ref{sec:honeypot} further considers reinforcement learning for  environmental uncertainties under the honeypot engagement scenario. The conclusion and discussion are presented in Section \ref{sec:conclusion}. 

\section{Bayesian Learning for Uncertain Parameters}
\label{sec:Type}
\begin{figure*}[h]
\sidecaption[t]
\centering
\includegraphics[width=0.63 \textwidth]{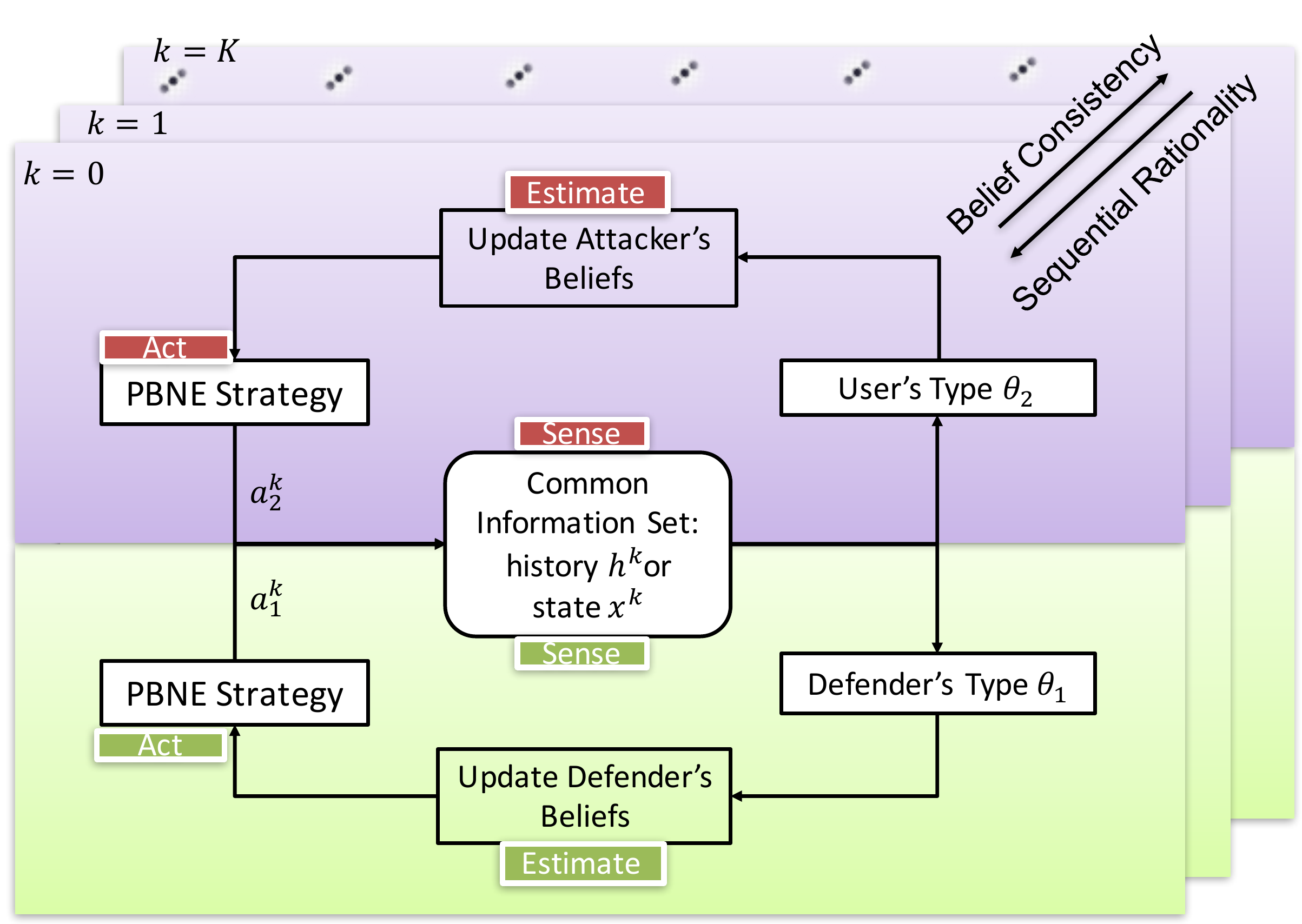}
\caption{
The feedback loop of the Bayesian learning from the initial stage $k=1$ to the terminal stage $k=K$. 
Each player forms a belief of the other player's type and persistently updates the belief based on the actions  resulted from the PBNE strategies which are the results of Bayesian learning. 
}
\label{fig: feedback1}
\end{figure*}

Under the mild restrictive information structure, each player' utility is completely governed by a finite group of parameters which form his/her \textit{type}. 
Each player's \textit{type} characterizes all the uncertainties about this player during the game interaction, e.g., the physical outcome, the payoff function, and the strategy feasibility, as an equivalent utility uncertainty without loss of generality  \cite{harsanyi1967games}. 
Thus, the revelation of the type value directly results in a game of complete information. 
In the cyber security scenario, a discrete type can distinguish either systems with different kinds of vulnerabilities or attackers with different targets. The type can also be a continuous random variable representing either the threat level or the security awareness level \cite{huang2019adaptive, huang2018analysis}. 
Since each player $P_i$ takes actions to maximize his/her own type-dependent utility, the other player $P_j$ can form a belief to estimate $P_i$'s type based on the observation of $P_i$'s action history.  
The utility optimization under the beliefs results in the Perfect Bayesian Nash Equilibrium (PBNE) which generates  new action samples and updates the belief via the Bayesian rule.  
We plot the feedback Bayesian learning process in Fig.  \ref{fig: feedback1} and elaborate each element in the following subsections based on our previous work \cite{APTjournal}. 

 \subsection{Type and Multistage Transition}
Through adversarial deception techniques, attackers can disguise their subversive actions as legitimate behaviors so that the defender $P_1$ cannot judge whether a user $P_2$'s type $\theta_2\in \Theta_2:=\{\theta_2^g,\theta_2^b\}$ is legitimate $\theta_2^g$ or adversarial $\theta_2^b$. 
As a countermeasure, the defender can introduce the defensive deception so that the attacker cannot distinguish between a primitive system  $\theta_1^L$ and a sophisticated system $\theta_1^H$, i.e., the defender has a binary type $\theta_1\in \Theta_1:=\{\theta_1^H,\theta_1^L\}$. 
A sophisticated system is costly yet deters attacks and causes damages to attackers. Thus, a primitive system can disguise as a sophisticated one to draw the same threat level to attackers yet avoid the implementation cost of sophisticated defense techniques. 

\begin{figure}
\centering
\includegraphics[width=1 \textwidth]{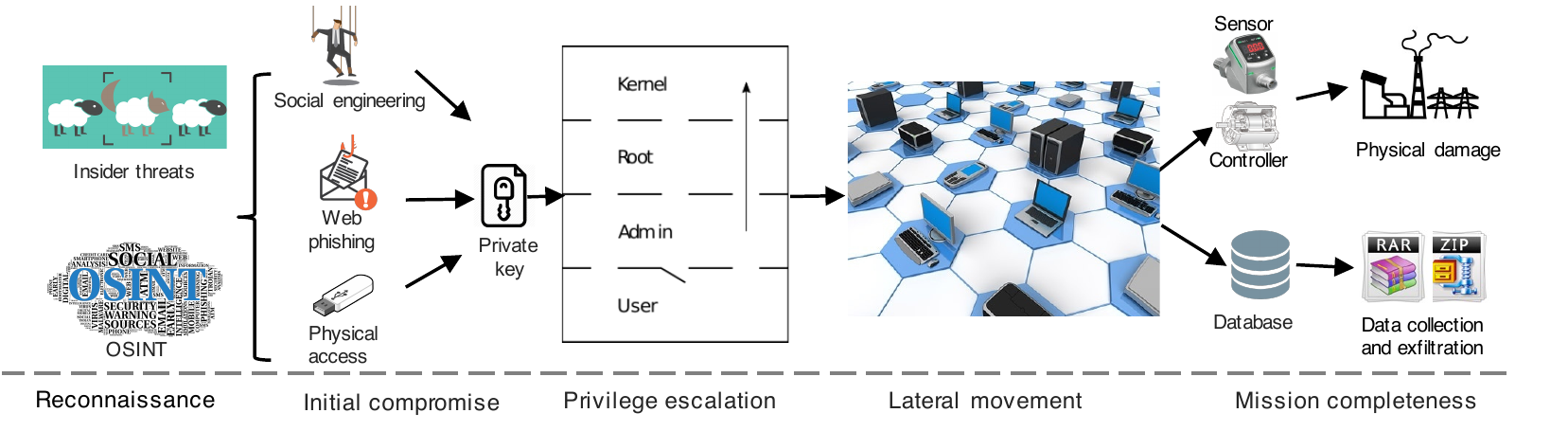}
\caption{
The multistage structure of APT kill chain is composed of reconnaissance, initial compromise, privilege escalation, lateral movement, and mission execution. 
}
 \label{fig: APT}
\end{figure}

Many cyber networks contain hierarchical layers, and up-to-date attackers such as Advanced Persistent Threats (APTs) aim to penetrate these layers and reach  specific targets at the final stage as shown in Fig. \ref{fig: APT}. 
%The superscript $k$ is the stage index. 

At stage $k\in \{0,1,\cdots,K\}$, $P_i$ takes an action $a_i^k\in \mathcal{A}_i^k$ from a finite and discrete set $\mathcal{A}_i^k$. 
Both players' actions become fully observable after applied and each action does not directly reveal the private type. 
For example, both legitimate and adversarial users can choose to access the sensor, and both primitive and sophisticated defenders can choose to monitor the sensor. 
%These behaviors are assumed to be fully observable by checking the log files. 
Both players' actions up to stage $k$ constitute the \textit{history} $h^k=\{a_1^0,\cdots,a_1^{k-1},a_2^0,\cdots,a_2^{k-1}\} \in \mathcal{H}^k:=\prod_{i=1}^2 \prod_{\bar{k}=0}^{k-1} \mathcal{A}_i^{\bar{k}}$. Given history $h^k$ at the current stage $k$, players at stage $k+1$ obtain an updated history
$h^{k+1}=h^k\cup \{a_1^k,a_2^k\}$ after the observation  $a_1^k,a_2^k$. 
A state $x^k\in \mathcal{X}^k$ at each stage $k$ is the smallest set of quantities that summarize information about actions in previous stages so that the initial state $x^0\in \mathcal{X}^0$ and the history at stage $k$ uniquely determine ${x}^k$ through a known state transition function $f^k$, i.e., $x^{k+1}=f^k(x^k,a_1^k,a_2^k), \forall k\in \{0,1,\cdots,K-1\}$. 
%no movement can also be regarded as an action
The state can represent the location of the user in the attack graph, and also other quantities such as users' privilege levels and status of sensor failures.

A behavioral strategy $\sigma^k_i \in \Sigma^k_i:\mathcal{I}^k_i \mapsto \bigtriangleup (\mathcal{A}_i^k) $ maps $P_i$'s information set $\mathcal{I}^k_i $ at stage $k$ to a probability distribution over the action space $\mathcal{A}_i^k$. 
At the initial stage $0$, since the only information available is the player's type realization, the information set $\mathcal{I}_i^0=\Theta_i$. 
The action is a realization of the behavioral strategy, or equivalently, a sample drawn from the probability distribution $\sigma^k_{i}(\cdot| I_i^k )$. 
With a slight abuse of notation, we denote $\sigma^k_i(a_i^k| {I}_i^k)$ as the probability of $P_i$ taking action $a_i^k\in \mathcal{A}_i^k$ given the available information $I_i^k\in \mathcal{I}_i^k$. 
%Note that the values of the other player's type $\theta_j$ and action $a_j^k$, which are not observable for $P_i$ at stage $k$, do not affect $P_i$'s behavior strategy $\sigma_i^k$, i.e., $\Pr({a}_i^k|{a}^k_{j},\theta_{j}, {I}_i^k)=\sigma_i^k({a}_i^k| {I}_i^k)$.  Thus, $\sigma_1^k$ and $\sigma_2^k$ are conditional independent, i.e., $\Pr({a}_i^k,{a}^k_{j}|{I}_i^k,{I}_j^k)=\sigma^k_i({a}_i^k| {I}_i^k)\sigma^k_{j}({a}^k_{j}| {I}_j^k)$. 

\subsection{Bayesian Update under Two Information Structure}
Since the other player's type is of private information, $P_i$ forms a belief $b^k_i: \mathcal{I}_i^k  \mapsto \bigtriangleup (\Theta_{j}), j\neq i$, on $P_j$'s type using the available information $\mathcal{I}^k_i$. 
Likewise,  given information $I_i^k\in \mathcal{I}_i^k$ at stage $k$, $P_i$ believes with a probability $b^k_i(\theta_j| {I}_i^k)$ that $P_j$ is of type $\theta_j\in \Theta_j$. 
The initial belief $b^0_i: \Theta_i \mapsto \bigtriangleup \Theta_{j}, \forall i,j\in \{1,2\},j\neq i$, is formed based 
on an imperfect detection, side-channel information or the statistic estimation resulted from past experiences. 
%If no previous experiences are available to $P_i$, $P_i$ can take the uniform distribution as an unbiased prior belief.  

%If we treat type as the parameter $Par$
%\begin{equation}
%\Pr(Par|Sam, M)=\frac{\Pr(Par|Sam, M)\times \Pr(Sam|Par, M)}{\Pr(Sam| M)}
%\end{equation}

If the system has a \textit{perfect recall} $\mathcal{I}_i^k=\mathcal{H}^k\times \Theta_i$, then players can update their beliefs according to the Bayesian rule: 
\begin{equation}
 {b}^{k+1}_i(\theta_{j}|h^{k}\cup \{a_i^k,a_j^k\}, \theta_i)=\frac{   \sigma^k_i({a}_i^k| {h}^k, \theta_i)\sigma^k_{j}({a}^k_{j}| {h}^k,\theta_{j}) b_i^k(\theta_{j}| {h}^k, \theta_i)    }
{\sum_{\bar{\theta}_{j}\in \Theta_j} \sigma^k_i({a}_i^k| {h}^k, \theta_i)\sigma^k_{j}({a}^k_{j}| {h}^k,\bar{\theta}_{j}) b_i^k(\bar{\theta}_{j}| {h}^k, \theta_i)   }.
\label{eq: history-dependent belief update}
\end{equation}
Here, $P_i$ updates the belief ${b}^{k}_i$ based on the observation of the action $a_i^k,a_j^k$. When the denominator is $0$, the history $h^{k+1}$ is not reachable from $h^k$, and a Bayesian update does not apply. In this case, we let 
 ${b}^{k+1}_i(\theta_{j}|h^{k}\cup \{a_i^k,a_j^k\}, \theta_i):=b_i^0(\theta_j |\theta_i)$. 

If the information set is taken to be $\mathcal{I}_i^k=\mathcal{X}^k\times \Theta_i$ with the Markov property that $\Pr(x^{k+1}| \theta_{j}, x^k, \cdots,x^1,x^0, \theta_i)=\Pr(x^{k+1}| \theta_{j}, x^k, \theta_i)$, then the  Bayesian update between two consequent states is 
\begin{equation}
\label{eq: state-dependent belief update}
b_i^{k+1}(\theta_{j} | x^{k+1},\theta_i)=\frac{\Pr(x^{k+1}| \theta_{j}, x^k,\theta_i)b_i^{k}(\theta_{j}|x^k,\theta_i)}{\sum_{\bar{\theta}_{j}\in \Theta_j} \Pr(x^{k+1}| \bar{\theta}_{j}, x^k,\theta_i)b_i^{k}(\bar{\theta}_{j}|x^k,\theta_i)}.
\end{equation}

%With the conditional independence of $\sigma^k_1$ and $\sigma^k_2$, %we can compute 
%\begin{equation}
%\label{eq: cluster}
%\begin{split}
%& \Pr(x^{k+1}| \theta_{j}, x^k,\theta_i)\\
%&=\sum_{\{a^k_1,a^k_2\}\in \mathcal{A}_{x^k}^{x^{k+1}} }\sigma^k_1(a^k_1|x^k,\theta_1)\sigma^k_2(a^k_2|x^k,\theta_2), 
%\end{split}
%\end{equation}
% where the set $\mathcal{A}_{x^k}^{x^{k+1}}:=\{a^k_1\in \mathcal{A}^k_1,a^k_2\in \mathcal{A}^k_2| x^{k+1}=f^k(x^k,a^k_1,a^k_2)\}$ contains the action pairs that change the system state from $x^k$ to $x^{k+1}$.  

%Equation \eqref{eq: cluster} shows that the Bayesian update in \eqref{eq: state-dependent belief update} can be obtained from \eqref{eq: history-dependent belief update} by clustering all the action pairs in the set $\mathcal{A}_{x^k}^{x^{k+1}}$. 
%that change the system state from $x^k$ to $x^{k+1}$. 
The Markov belief update \eqref{eq: state-dependent belief update} can be regarded as an approximation of \eqref{eq: history-dependent belief update} using action aggregations. 
Unlike the history set $\mathcal{H}^k$, the dimension of the state set $|\mathcal{X}^k|$ does not grow with the number of stages. Hence, the Markov approximation significantly reduces the
 memory and computational complexity. 
 
%The following sections adopt the Markov belief update in \eqref{eq: state-dependent belief update}. For every given state $x^k\in \mathcal{X}^k$, we can use a type-dependent probability vector $Q_i^k(\theta_i):=[\sigma_i^k(a_i^{i,k} | x^k, \theta_i), \cdots, \sigma_i^k(a_i^{m_i^k,k} | x^k, \theta_i)]' \in \Gamma^{m_i^k}$ to represent the behavioral strategy $\sigma_i^k(\cdot | x^k, \theta_i)$. 
%\begin{equation*}
%P(\theta_{-i}, {h}^t,{a}_i^t,{a}^t_{-i}, \theta_i)=P({a}_i^t,{a}^t_{-i}|\theta_{-i}, {h}^t, \theta_i)P(\theta_{-i}, {h}^t, \theta_i)\\
%=P({a}_i^t|\theta_{-i}, {h}^t, \theta_i)P({a}^t_{-i}|\theta_{-i}, {h}^t, \theta_i)P(\theta_{-i}, {h}^t, \theta_i)\\
%=P({a}_i^t| {h}^t, \theta_i)P({a}^t_{-i}|\theta_{-i}, {h}^t)P(\theta_{-i}| {h}^t, \theta_i)P({h}^t, \theta_i)
%\end{equation*}
%$P({h}^t, \theta_i)$ is known at stage $t$, thus with probability 1. 
%To prove the conditional independency, we only need to prove that $P(X|Y,Z)=P(X|Z)$, where $X=a_i, Y=a_{-i}$ and $Z=\{h, \theta_i,\theta_{-i}\}$. 
%However, by doing the above computation, we are not be able to obtain Markov belief, i.e., we will get history-dependent(the dimension grows so quickly with stage $3*3\times 2*2\times 2*2=144$ different belief for the final stage) rather than state-dependent belief. 
%So we have to use the alternative form to compute the Markov belief as follows. 

\subsection{Utility and PBNE}
\label{subsec: utility}
At each stage $k$, $P_i$'s stage utility $\bar{J}_i^k: \mathcal{X}^k \times \mathcal{A}_1^k \times \mathcal{A}_2^k \times \theta_1 \times \theta_2 \times \mathcal{R} \mapsto \mathcal{R} $ depends on both players' types and actions, the current state ${x}^k\in \mathcal{X}^k$, and an external noise $w_i^k\in \mathcal{R}$ with a known probability density function $\varpi_i^k$. 
The noise term models unknown or uncontrolled factors that can affect the value of the stage utility. 
%The existence of the external noise makes it impossible for $P_i$, after reaching stage $k+1$,  to infer the value of the other player's type $\theta_j$ based on the knowledge of the input parameters $x^k , a^k_1 , a^k_2 , \theta_i$,  together with the output of the utility function $\bar{J}_i^k$ at stage $k$. 
Denote the expected stage utility as $J_i^k(x^k, a^k_1 , a^k_2 , \theta_1,\theta_2):=E_{w_i^k\sim \varpi_i^k }\bar{J}_i^k(x^k, a^k_1 , a^k_2 , \theta_1,\theta_2,w_i^k), \forall x^k, a^k_1 , a^k_2 , \theta_1,\theta_2$. 

%We use the privilege escalation example at state $x^k$ of stage $k$ in Table \ref{table: muti-stageBay} to show that the same action taken by $P_i$ of different types can lead to different utilities $J_i^k(x^k, a^k_1 , a^k_2 , \theta_1,\theta_2)$.  

%\begin{table}[h]
%\sidecaption
%\centering
%\caption{Utility bi-matrix $(J_1^k, J_2^k)$ for users and defenders of %different types. 
%}
%\label{table: muti-stageBay}
%\begin{tabular}{|l|l|l|l|}
%\hline 
%$\theta_2=\theta^b$  &   NOP & Escalate\\ \hline
%Permit               & $(0,0)$   & $(-r_2, r_2)$    \\ \hline
%Restrict          & $(0,0)$  & $(r,-r)$   \\ \hline
%\end{tabular}
%\quad
%\begin{tabular}{|l|l|l|l|}
%\hline 
%$\theta_2=\theta^g$   &   NOP & Escalate\\ \hline
%Permit               & $(0,0)$   & $(r_1, r_1)$    \\ \hline
%Restrict          & $(0,0)$  & $(-r_1,-r_1)$   \\ \hline
%\end{tabular}
%\end{table}

%If the defender permits the privilege escalation from a legitimate user $\theta_2=\theta_2^g$, then both players receive a reward $r_1$. 
%If the defender permits an escalation request from an attacker, then the attacker obtains reward $r_2$ and the defender losses $r_2$. 
%If the defender chooses to restrict every escalation requests, then legitimate user suffers a loss of $r_1$ and the attacker suffers a loss of $r:=r_L \cdot \mathbf{1}_{\{\theta_1=\theta_1^L\}}+r_H \cdot \mathbf{1}_{\{\theta_1=\theta_1^H\}}$. The loss is slight under a primitive system yet considerable under a sophisticated system, i.e., $r_H>r_L$. 

Given the  type $\theta_i\in \Theta_i$, the initial state $x^{k_0}\in \mathcal{X}^{k_0}$, and both players' strategies $\sigma_i^{k_0:K}:=[\sigma^k_i(a_i^k|{x}^{k},\theta_i)]_{k=k_0,\cdots,K}\in \prod_{k=k_0}^K \Sigma_i^k$ from stage $k_0$ to $K$, we can determine the expected cumulative utility $U_i^{k_0:K}$ for $P_i, i\in \{1,2\}$, by taking expectations over the mixed-strategy distributions and the $P_i$'s belief on $P_j$'s type, i.e., 
\begin{equation}
\label{eq: cumultive utility}
U^{k_0:K}_i(\sigma_i^{k_0:K},\sigma_{j}^{k_0:K}, x^{k_0},\theta_i) :=\sum_{k=k_0}^{K} E_{\theta_{j}\sim b_i^k, a_i^k\sim \sigma_i^k,a_j^k \sim \sigma_{j}^k} J_i^k({x}^k,a_1^k,a_{2}^k,\theta_1,\theta_{2}).
\end{equation}

The attacker and the defender use the Bayesian update to reduce their uncertainties on the other player's type.  Since their actions affect the belief update, both players at each stage should optimize their expected cumulative utilities concerning 
%long-term strategies based on the current utility and the expected future utility under 
the updated beliefs, which leads to the solution concept of PBNE in Definition \ref{def: PBNE}. 
%For the multistage game of incomplete information defined in Section \ref{sec:model}, Definition \ref{def: PBNE} states the solution concept of PBNE. 

%[BEFORE PBNE, STATE THAT PBNE IS THE APPROPRIATE SOLUTION CONCEPT THAT WE ARE LOOKING FOR IN THE CONTEXT OF OUR APPLICATIONS.]

\begin{definition}
\label{def: PBNE}
Consider the two-person $K$-stage game with a double-sided incomplete information, a sequence of beliefs $b_i^k, \forall k\in \{0,\cdots, K\}$, an expected cumulative utility $U^{0:K}_i$ in \eqref{eq: cumultive utility}, and a given scalar $\varepsilon\geq 0$. A sequence of strategies $\sigma_i^{*,0:K}\in \prod_{k=0}^K \Sigma_i^k$ is called 
$\varepsilon$-perfect Bayesian Nash equilibrium for player $i$  if the following two conditions are satisfied. 
\begin{itemize}
\item[C1]: Belief consistency: under the strategy pair $(\sigma_1^{*,0:K},\sigma_2^{*,0:K})$, each player's belief $b_i^{k}$ at each stage $k=0,\cdots, K$ satisfies %the Bayesian belief update in
\eqref{eq: state-dependent belief update}. 
\item[C2]: Sequential rationality: for all given initial state $x^{k_0}\in \mathcal{X}^{k_0}$ at every initial stage $k_0\in \{0,\cdots,K\}$, $\forall \sigma_{1}^{k_0:K}\in \prod_{k=0}^K \Sigma_1^k, \forall \sigma_{2}^{k_0:K}\in \prod_{k=0}^K \Sigma_2^k$, 
\begin{equation}
\label{eq: PBNE in def}
\begin{split}
&U_1^{k_0:K}(\sigma_1^{*,k_0:K},\sigma_{2}^{*,k_0:K}, {x}^{k_0},\theta_1)+\varepsilon
\geq 
U_1^{k:K}(\sigma_1^{k_0:K},\sigma_{2}^{*,k_0:K}, {x}^{k_0},\theta_1),   
\\
&U_2^{k_0:K}(\sigma_1^{*,k_0:K},\sigma_{2}^{*,k_0:K}, {x}^{k_0},\theta_2)+\varepsilon
\geq  
U_2^{k:K}(\sigma_1^{*,k_0:K},\sigma_{2}^{k_0:K}, {x}^{k_0},\theta_2).  
\end{split}
\end{equation}
\end{itemize}
When  $\varepsilon = 0$, the equilibrium is called  Perfect Bayesian Nash Equilibrium (\textbf{PBNE}). 
\qed
\end{definition}

Solving PBNE is challenging. 
If the type space is discrete and finite, then given each player's belief at all stages, we can solve the equilibrium strategy satisfying condition C2 via dynamic programming and a bilinear program. Next, we update the belief at each stage based on the computed equilibrium strategy. We iterate the above update on the equilibrium strategy and belief until they satisfy condition C1 as demonstrated in \cite{APTjournal}. 
If the type space is continuous, then the Bayesian update can be simplified into a parametric update under the conjugate prior assumption. Next, the parameter after each belief update can be assimilated into the backward dynamic programming of equilibrium strategy with an expanded state space \cite{huang2018analysis}. 
Although no iterations are required, the infinite dimension of continuous type space limits the computation to two by two game matrices. 

\begin{figure}[h]
\sidecaption[t]
\includegraphics[width=0.6  \textwidth]{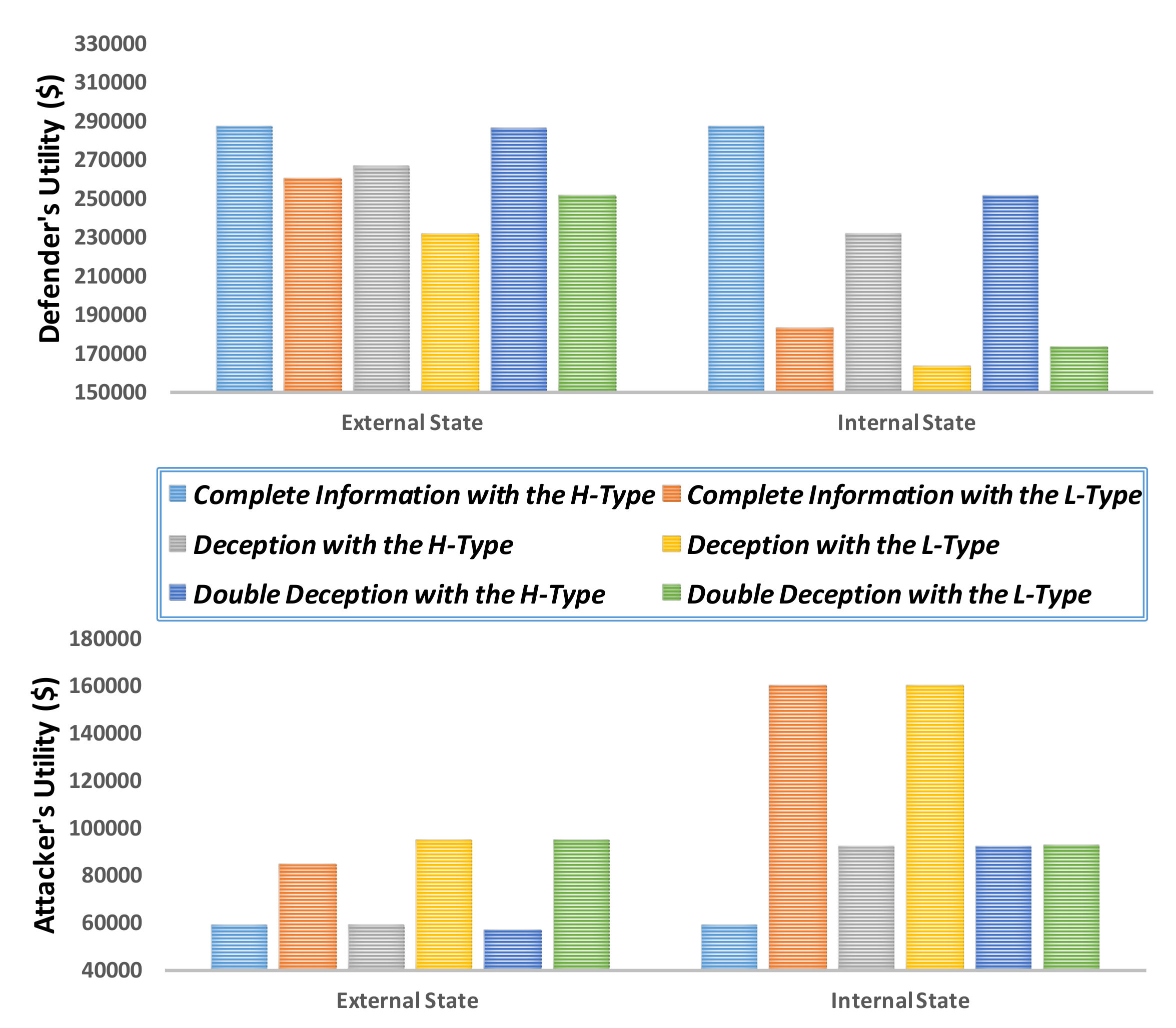}
\caption{
The cumulative utilities of the attacker and the defender under the complete information, the adversarial deception, and the defensive deception.  
%In the legend, the left three represent the utilities for a system of type $\theta_1^H$ and the right three represent the ones for a system of type $\theta_1^L$. 
The complete information refers to the scenario where both players know the other player's type. 
The deception with the H-type or the L-type means that the attacker knows the defender's type to be $\theta_1^H$ or $\theta_1^L$, respectively,  yet the defender has no information about the user's type. 
The double-sided deception indicates that both players do not know the other player's type. 
}
 \label{fig: v12compare}
\end{figure}

We apply the above framework and analysis to a case study of Tennessee Eastman (TE) process and  investigate both players' multistage utilities under the adversarial and the defensive deception in Fig. \ref{fig: v12compare}. Some insights are listed as follows. 
%\begin{figure*}[t]
%\minipage{0.32\textwidth}
%  \includegraphics[width=1 \textwidth]{Fig/Finalstagebelief1Revise2}
%\caption{
%The SBNE strategy and the expected utility when the defender has one-sided incomplete information. %under a different percentage of adversarial users. 
% \label{fig: FinalstageBlief}}
%\endminipage\hfill
%\minipage{0.32\textwidth}
%  \includegraphics[width=1 \textwidth]{Fig/Finalstagebelief2revise2}
%\caption{
%The SBNE strategy and the expected utility when the user has one-sided incomplete information. 
% \label{fig: Finalstagebelief2revise}}
%\endminipage\hfill
%\minipage{0.32\textwidth}%
%  \includegraphics[width=1 \textwidth]{Fig/Finalstagebelief4revise2}
%\caption{
% \label{fig: Finalstagebelief3revise}
% The SBNE strategy and the expected utility under double-sided incomplete information.  %The user knows with a probability $0.5$ that the system is advanced. 
%}
%\endminipage
%\end{figure*}

%The results from Fig. \ref{fig: v12compare} are summarized as follows. 
First, the defender's payoffs under type $\theta_1^H$ can increase as much as $56\%$ than those under type $\theta_1^L$. 
% Also, prevention of the attacker from entering the internal network $x^0=1$ increases the defender's utility by as much as $41\%$ and reduces the attacker's utility  by as much as $38\%$. 
Second, the defender and the attacker receive the highest and the lowest payoff, respectively, under the complete information.
%\bf{[IS ATTACKER P2? P1 IS THE SYSTEM? IS IT NECESSARY TO USE THE LINGO P1 AND P2 OR JUST STAY ATTACKER AND THE DEFENDER]} 
When the attacker introduces deceptions over his type, the attacker's utility increases and the system utility decreases. 
Third, when the defender adopts defensive deceptions to introduce double-sided incomplete information, we find that the decrease of system utilities is reduced by at most $64 \%$, i.e., the decrease of system utilities changes from $\$55,570$ to $\$35,570$ under the internal state and type $\theta_1^H$.  
The double-sided incomplete information also brings lower utilities to the attacker than the one-sided adversarial deception. 
However, the system utility under the double-sided deception is still less than the complete information case, which concludes that acquiring complete information of the adversarial user is the most effective defense. However, if the complete information cannot be obtained, the defender can mitigate her loss by introducing defensive deceptions. 

\section{Distributed Learning for Uncertain Payoffs}
\label{sec:MTD}
In the previous section, we study known attacks and systems that  adopt cyber deception to conceal their types. 
We assume common knowledge of the prior probability distribution of the unknown type, and also a common observation of either the action history or the state at each stage. 
Thus, each player can use Bayesian learning to reduce the  other player's type uncertainty. 

In this section, we consider unknown attacks in the MTD game stated in \cite{zhu2013game} 
where each player has no information on the past actions of the other player, and the payoff functions are subject to noises and disturbances with unknown statistical characteristics.  
Without information sharing between players, the learning is distributed. 

\subsection{Static Game Model of MTD}
We consider a system of $N$ layers yet focus on the  static game at layer $l\in \{1,2,\cdots,N\}$ because the technique can be employed at each layer of the system independently. 
At layer $l$, $\mathcal{V}_l:=\{v_{l,1},v_{l,2},\cdots,v_{l,n_l}\}$ is the set of $n_l$ system vulnerabilities that an attacker can exploit to compromise the system. 
Instead of a static configuration at layer $l$, the defender can choose to change her configuration from a finite set of $m_l$ feasible configurations $\mathcal{C}_l:=\{c_{l,1},c_{l,2},\cdots,c_{l,m_l}\}$. 
Different configurations result in different subsets of vulnerabilities among $\mathcal{V}_l$, which are characterized by the vulnerability map $\pi_l:\mathcal{C}_l \rightarrow 2^{\mathcal{V}_l}$. 
We call $\pi_l(c_{l,j})$ the attack surface at stage $l$ under configuration $c_{l,j}$. 

Suppose that for each vulnerability $v_{l,j}$, the attacker can take a corresponding attack $a_{l,j}=\gamma_l(v_{l,j})$ from the action set $\mathcal{A}_l:=\{a_{l,1},a_{l,2},\cdots,a_{l,n_l}\}$. Attack action $a_{l,j}$ is only effective and incurs a bounded cost $D_{ij}\in \mathbb{R}_{+}$ when the vulnerability $v_{l,j}=\gamma_l^{-1}(a_{l,j})$ exists in the current attack surface $ \pi_l(c_{l,k})$. 
Thus, the damage caused by the attacker at stage $l$ can be represented as 
 
  \begin{equation}
    r_l(a_{l,j},c_{l,i})=
    \begin{cases}
      D_{ij}, &  \gamma_l^{-1}(a_{l,j})\in \pi_l(c_{l,k}) \\
      0, & \text{otherwise}
    \end{cases}.  
  \end{equation}
%where $D_{ij}\in \mathbb{R}_{+}$ is the bounded damage quantified in terms of monetary values. 

Since vulnerabilities are inevitable in a modern computing system, we can randomize the configuration and make it difficult for the attacker to learn and locate the system vulnerability, which naturally leads to the mixed strategy equilibrium solution concept of the game. 
At layer $l$, the defender's strategy $\mathbf{f}_l=\{f_{l,1},f_{l,2},\cdots,f_{l,m_l}\}\in \bigtriangleup \mathcal{C}_l$ assigns probability $f_{l,j}\in [0,1]$ to configuration $c_{l,j}$ while the attacker's strategy $\mathbf{g}_l:=\{g_{l,1},g_{l,2},\cdots,g_{l,n_l}\}\in \bigtriangleup \mathcal{A}_l$ assigns probability $g_{l,i}\in [0,1]$ to attack action $a_{l,i}$. 
The zero-sum game possesses a mixed strategy saddle-point equilibrium (SPE) $(\mathbf{f}_l^*\in  \bigtriangleup \mathcal{C}_l,\mathbf{g}_l^*\in  \bigtriangleup \mathcal{A}_l),$ and a unique game value $\mathbbm{r}(\mathbf{f}_l^*,\mathbf{g}_l^*)$, i.e., 
\begin{equation}
\mathbbm{r}_l(\mathbf{f}_l^*,\mathbf{g}_l)\leq \mathbbm{r}_l(\mathbf{f}_l^*,\mathbf{g}_l^*) \leq \mathbbm{r}_l(\mathbf{f}_l,\mathbf{g}_l^*), \forall \mathbf{f}_l \in  \bigtriangleup \mathcal{C}_l, \mathbf{g}_l \in  \bigtriangleup \mathcal{A}_l, 
\end{equation}
where the expected cost $\mathbbm{r}_l$ is given by
\begin{equation}
\mathbbm{r}_l(\mathbf{f}_l,\mathbf{g}_l) :=\mathbb{E}_{\mathbf{f}_l,\mathbf{g}_l}r_l =\sum_{k=1}^{n_l}\sum_{h=1}^{m_l}f_{l,h}g_{l,k}r_l(a_{l,k},c_{l,h}). 
\end{equation}

\begin{figure}[h]
    \centering
    \begin{subfigure}[]{0.45\textwidth}
        \centering
        \includegraphics[width=1 \textwidth]{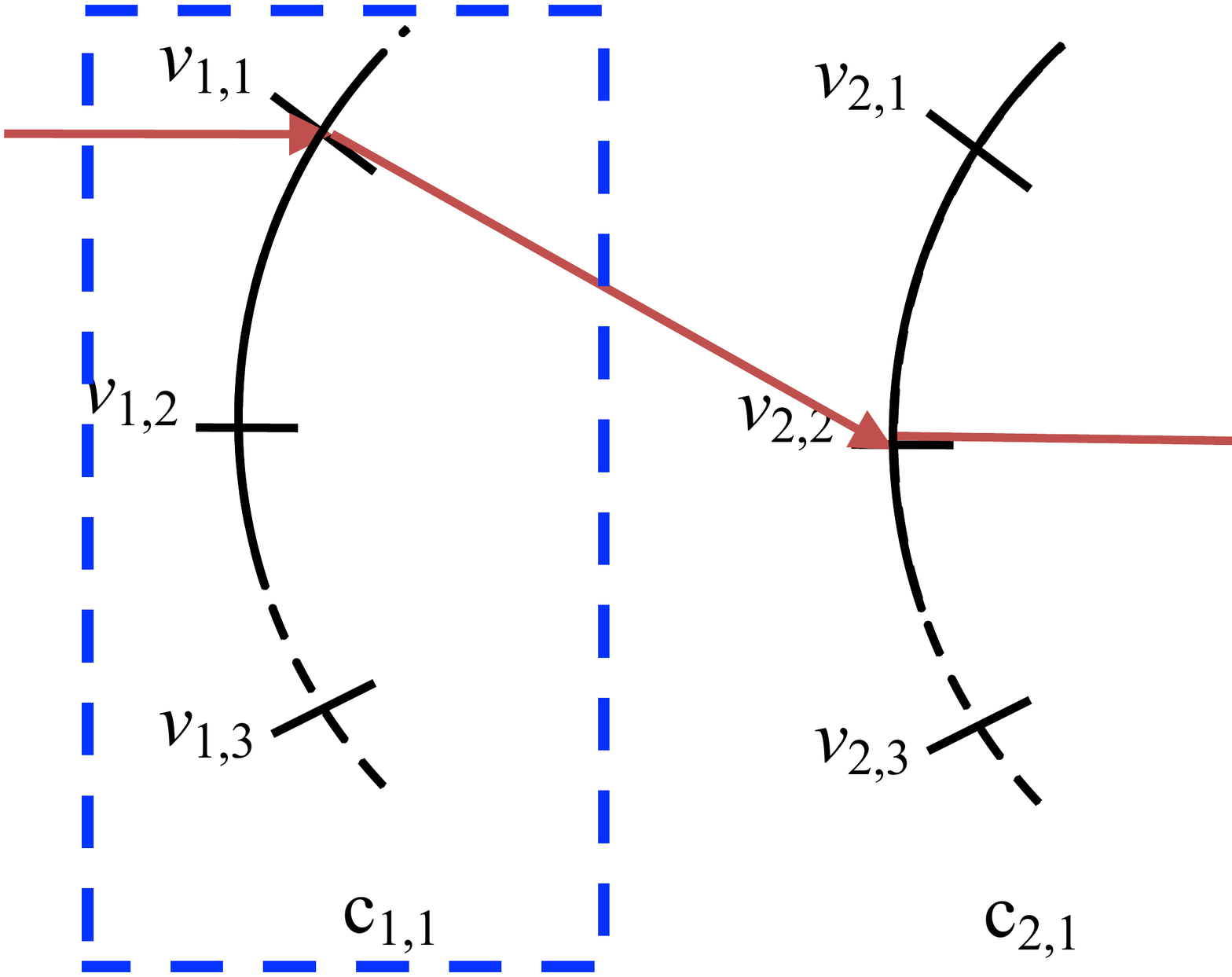}
        \caption{Attack surface $\pi_1(c_{1,1})=\{v_{1,1},v_{1,2}\}$. }
        \label{fig:config1}
    \end{subfigure}%
    \hfill 
    \begin{subfigure}[]{0.45\textwidth}
        \centering
        \includegraphics[width=1 \textwidth]{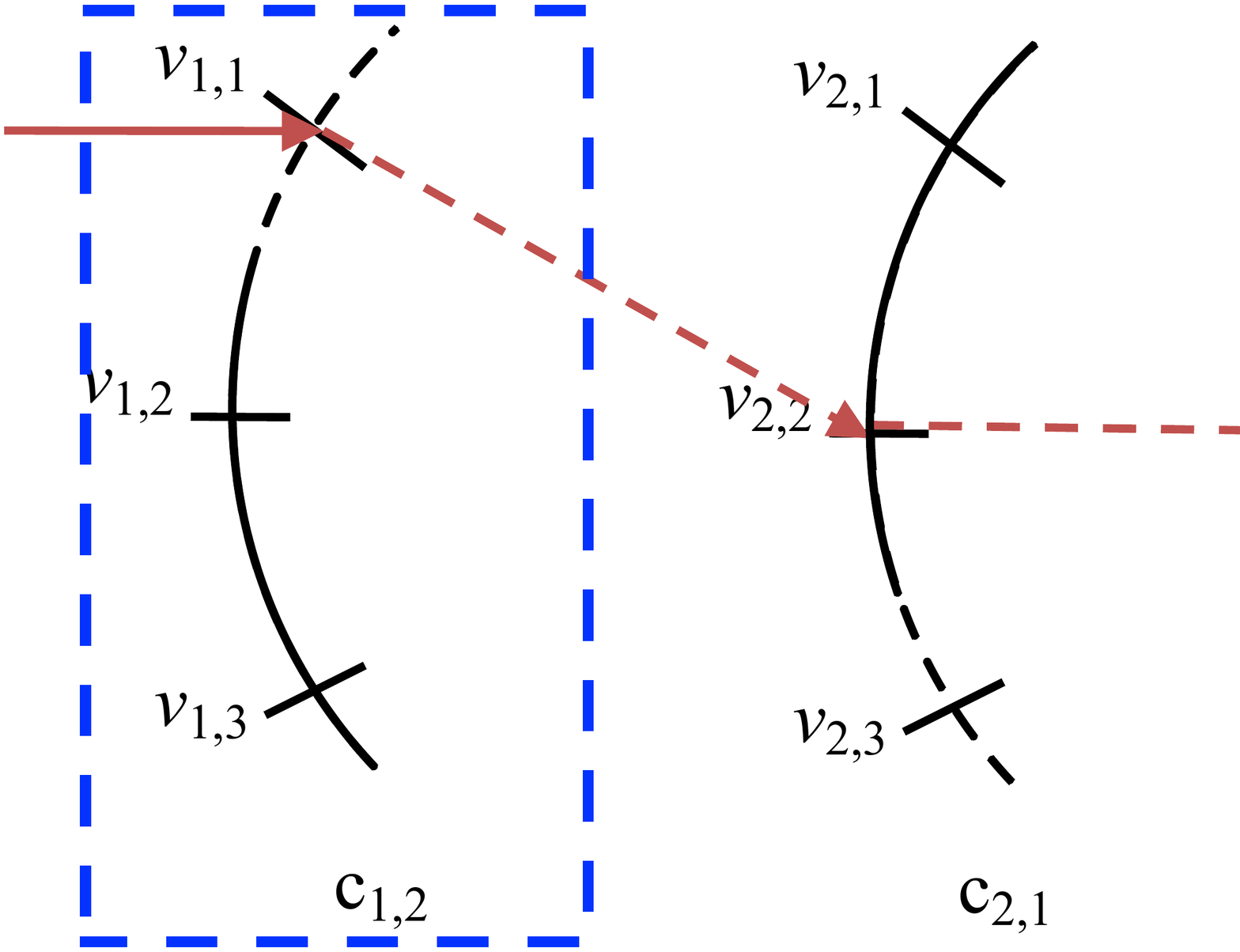}
        \caption{Attack surface $\pi_1(c_{1,2})=\{v_{1,2},v_{1,3}\}$. }
         \label{fig:config2}
    \end{subfigure}
    \caption{
    Given a static configuration $c_{1,1}$, an attacker can succeed in reaching the resources at deeper layers by forming an attack path $v_{1,1}\rightarrow v_{2,2}\rightarrow \cdots$. A change of configuration to $c_{1,2}$ can thwart the attacker at the first layer. 
 }
  \label{fig:config12}
\end{figure}

We illustrate the multistage MTD game in Fig. \ref{fig:config12} and focus on the first layer with two available configurations $\mathcal{C}_1:=\{c_{1,1},c_{1,2}\}$
 in the blue box. 
Configuration $c_{1,1}$  in Fig. \ref{fig:config1}  has an attack surface $\pi_1(c_{1,1})=\{v_{1,1},v_{1,2}\}$ while configuration $c_{1,2}$ in Fig. \ref{fig:config2} reveals two vulnerabilities $v_{1,2},v_{1,3}\in \pi_1(c_{1,2}) $. 
Then, if the attacker takes action $a_{1,1}$ and the defender changes the configuration from $c_{1,1}$ to $c_{1,2}$, the attack is deterred at the first layer. 

\subsection{Distributed Learning}
In practical cybersecurity domain, the payoff function $r_l$ is subjected to noises of unknown distributions. 
Then, each player reduces the payoff uncertainty by repeatedly observing the payoff realizations during the  interaction with the other player. 
We use subscript $t$ to denote the strategy or cost at time $t$.

\begin{figure*}[h]
\sidecaption[t]
\centering
\includegraphics[width=0.64 \textwidth]{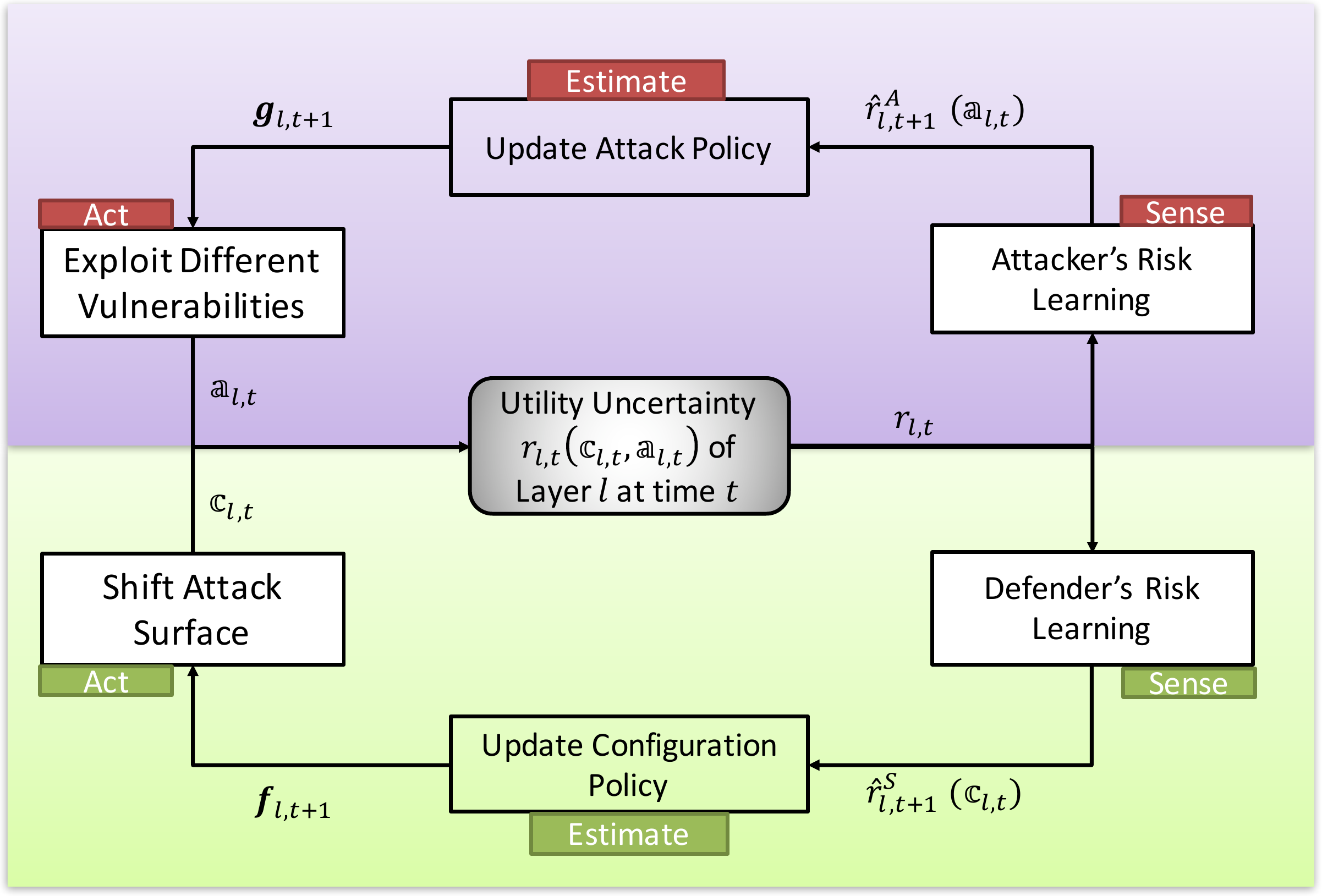}
\caption{ 
The distributed learning of the multistage MTD game at layer $l$. Adversarial learning in red does not share information with defensive learning in green.  
The distributed learning fashion means that the learning rule does not depend on the other player's action, yet the observed payoff depends on both players' actions.  
}
\label{fig: Feedback2}
\end{figure*}

There is no communication at any time between two agents due to the non-cooperative environment, and the configuration and attack action are kept private, i.e., each player cannot observe the other player's action. 
Thus, each player independently chooses action $\mathbbm{c}_{l,t}\in \mathcal{C}_l$ or $\mathbbm{a}_{l,t}\in \mathcal{A}_l$ to 
estimate the average risk of the system $\hat{r}_{l,t}^S: \mathcal{C}_l \rightarrow \mathbb{R}_{+}$ and $\hat{r}_{l,t}^A: \mathcal{A}_l \rightarrow \mathbb{R}_{+}$ at layer $l$. 
Based on the estimated average risk $\hat{r}_{l,t}^S$ and the previous policy $\mathbf{f}_{l,t}$, the defender can obtain her updated policy $\mathbf{f}_{l,t+1}$. 
Likewise, the attacker can also update his policy $\mathbf{g}_{l,t+1}$ based on $\hat{r}_{l,t}^A$ and $\mathbf{g}_{l,t}$. 
The new policy pair $(\mathbf{f}_{l,t+1},\mathbf{g}_{l,t+1})$ determines the next payoff sample. 
The entire distributed learning feedback loop is illustrated in  Fig. \ref{fig: Feedback2} where we distinguish the adversarial and defensive learning in red and green, respectively. 

In particular, players update their estimated average risks  based on the payoff sample $r_{l,t}$ under the chosen action pair $(\mathbbm{c}_{l,t},\mathbbm{a}_{l,t})$ as follows. Let $\mu_t^S$ and $\mu_t^A$ be the payoff learning rate for the system and attacker, respectively. 
\begin{equation}
\label{eq:utilityLearning}
\begin{split}
& \hat{r}_{l,t+1}^S(c_{l,h})=  \hat{r}_{l,t}^S(c_{l,h}) +\mu_t^S \mathbf{1}_{\{\mathbbm{c}_{l,t}=c_{l,h}\}} (r_{l,t}-\hat{r}_{l,t}^S(c_{l,h})), \\
& \hat{r}_{l,t+1}^A(a_{l,h})=\hat{r}_{l,t}^A(a_{l,h}) +\mu_t^A \mathbf{1}_{\{\mathbbm{a}_{l,t}=a_{l,h}\}} (r_{l,t}-\hat{r}_{l,t}^A(a_{l,h})). 
\end{split}
\end{equation}
The indicators in \eqref{eq:utilityLearning} mean that both players only update the estimate average risk of the current action. 

\subsubsection{Security versus Usability}
Frequent configuration changes may achieve the complete security yet also decrease the system usability. 
To quantify the tradeoff between the security and the  usability, we introduce the switching cost of policy from $\mathbf{f}_{l,t}$ to $\mathbf{f}_{l,t+1}$ as their entropy: 
\begin{equation}
R_{l,t}^S:=\sum_{h=1}^{m_l}f_{l,h,t+1}\ln\left(\frac{f_{l,h,t+1}}{f_{l,h,t}}\right). 
\end{equation} 
Then, the total cost at time $t$ combines the expected cost with the entropy penalty in a ratio of $\epsilon_{l,t}^S$. When $\epsilon_{l,t}^S$ is high, the policy changes less and is more usable, yet may cause a large loss and be less rational. 
\begin{equation}
\label{eq:SP}
(\texttt{SP}): \sup_{\mathbf{f}_{l,t+1}\in \bigtriangleup \mathcal{C}_l} -\sum_{h=1}^{m_l} f_{l,h,t+1} \hat{r}^S_{l,t} (c_{l,h})-\epsilon_{l,t}^S R_{l,t}^S. 
\end{equation}

A similar learning cost is introduced for the attacker:  
\begin{equation}
\label{eq:AP}
(\texttt{AP}):  \sup_{\mathbf{g}_{l,t+1}\in \bigtriangleup \mathcal{A}_l} -\sum_{h=1}^{n_l} g_{l,h,t+1} \hat{r}^A_{l,t} (a_{l,h})-\epsilon_{l,t}^A \sum_{h=1}^{n_l}g_{l,h,t+1}\ln\left(\frac{g_{l,h,t+1}}{g_{l,h,t}}\right). 
\end{equation}

At any time $t+1$, we are able to obtain the equilibrium strategy $(f_{l,h,t+1},g_{l,h,t+1})$ and game value $(W_{l,t}^S,W_{l,t}^A)$ in closed form of the previous strategy and the estimated average risk at time $t$ 
as follows.  
\begin{equation}
\label{eq:closedform}
\begin{split}
& f_{l,h,t+1}=\ddfrac{f_{l,h,t}  e^{ -\frac{\hat{r}_{l,t} (c_{l,h})}{\epsilon_{l,t}^S} } }
{\sum_{h'=1}^{m_l} f_{l,h',t} e^{ -\frac{\hat{r}_{l,t} (c_{l,h'})}{\epsilon_{l,t}^S} }}, \quad \quad
g_{l,h,t+1}=\ddfrac{g_{l,h,t}  e^{ -\frac{\hat{r}_{l,t} (a_{l,h})}{\epsilon_{l,t}^A} } }
{\sum_{h'=1}^{n_l} g_{l,h',t} e^{ -\frac{\hat{r}_{l,t} (a_{l,h'})}{\epsilon_{l,t}^A} }},  \\
& W_{l,t}^S=\epsilon_{l,t}^S \ln \left(\sum_{h=1}^{m_l}  f_{l,h,t}  e^{ -\frac{\hat{r}_{l,t} (c_{l,h})}{\epsilon_{l,t}^S} }  \right) , \quad \quad 
W_{l,t}^A=\epsilon_{l,t}^A \ln \left(\sum_{h=1}^{n_l}  g_{l,h,t}  e^{ -\frac{\hat{r}_{l,t} (a_{l,h})}{\epsilon_{l,t}^A} } \right).  
\end{split}
\end{equation}

\subsubsection{Learning Dynamics and ODE Counterparts}
The closed form of policy leads to the following learning dynamics with learning rates $\lambda_{l,t}^S,\lambda_{l,t}^A\in [0,1]$. 
\begin{equation}
\label{eq:policyLearning}
\begin{split}
f_{l,h,t+1}=(1-\lambda_{l,t}^S)f_{l,h,t}+\lambda_{l,t}^S\ddfrac{f_{l,h,t}  e^{ -\frac{\hat{r}_{l,t} (c_{l,h})}{\epsilon_{l,t}^S} } }
{\sum_{h'=1}^{m_l} f_{l,h',t} e^{ -\frac{\hat{r}_{l,t} (c_{l,h'})}{\epsilon_{l,t}^S} }}, \\
g_{l,h,t+1}=(1-\lambda_{l,t}^A)g_{l,h,t}+\lambda_{l,t}^A \ddfrac{g_{l,h,t}  e^{ -\frac{\hat{r}_{l,t} (a_{l,h})}{\epsilon_{l,t}^A} } }
{\sum_{h'=1}^{n_l} g_{l,h',t} e^{ -\frac{\hat{r}_{l,t} (a_{l,h'})}{\epsilon_{l,t}^A} }}. 
\end{split}
\end{equation}
If $\lambda_{l,t}^S=1,\lambda_{l,t}^A=1$, \eqref{eq:policyLearning} is the same as \eqref{eq:closedform}. 
According to the stochastic approximation theory, the convergence of the policy and the average risk requires the learning rates $\lambda_{l,t}^A,\lambda_{l,t}^S,\mu_{l,t}^A,\mu_{l,t}^S$ to satisfy the regular condition of convergency in Definition \ref{def:convergencyCondition}. 
\begin{definition}
\label{def:convergencyCondition}
A number sequence $\{x_t\}, t=1,2,\cdots$, is said to satisfy the regular condition of convergency if 
\begin{equation}
\sum_{t=1}^{\infty } x_t=+\infty, \quad \sum_{t=1}^{\infty } (x_t)^2<+\infty. 
\end{equation}
\qed
\end{definition}
The coupled dynamics of the payoff learning \eqref{eq:utilityLearning} and policy learning\eqref{eq:policyLearning} converge to their Ordinary Differential Equations (ODEs) counterparts in system dynamics \eqref{eq:systemDynaimcs} and attacker dynamics \eqref{eq:attackDynaimcs}, respectively. 
Let $e_{c_{l,h}}\in \bigtriangleup \mathcal{C}_l, e_{a_{l,h}}\in \bigtriangleup \mathcal{A}_l$ be vectors of proper dimensions with the $h$-th entry being $1$ and others being $0$. 
\begin{equation}
\label{eq:systemDynaimcs}
\begin{split}
&\frac{d}{dt} f_{l,h,t}=f_{l,h,t}\left(\ddfrac{ e^{ -\frac{\hat{r}_{l,t} (c_{l,h})}{\epsilon_{l,t}^S} } }
{\sum_{h'=1}^{m_l} f_{l,h',t} e^{ -\frac{\hat{r}_{l,t} (c_{l,h'})}{\epsilon_{l,t}^S} }}-1 \right), h=1,2,\cdots,m_l, \\
&\frac{d}{dt}  \hat{r}_{l,t}^S(c_{l,h})=  -\mathbbm{r}_{l,t}(e_{c_{l,h}}, \mathbf{g}_{l,t})-\hat{r}_{l,t+1}^S(c_{l,h}),  c_{l,h}\in \mathcal{C}_l. 
\end{split}
\end{equation}
\begin{equation}
\label{eq:attackDynaimcs}
\begin{split}
&\frac{d}{dt}   g_{l,h,t+1}= g_{l,h,t}  \left( \ddfrac{ e^{ -\frac{\hat{r}_{l,t} (a_{l,h})}{\epsilon_{l,t}^A} } }
{\sum_{h'=1}^{n_l} g_{l,h',t} e^{ -\frac{\hat{r}_{l,t} (a_{l,h'})}{\epsilon_{l,t}^A} }}-1 \right) , h=1,2,\cdots,n_l, \\
&\frac{d}{dt}  \hat{r}_{l,t+1}^A(a_{l,h})=\mathbbm{r}_{l,t}( \mathbf{f}_{l,t},e_{a_{l,h}})-\hat{r}_{l,t+1}^A(a_{l,h}),  a_{l,h}\in \mathcal{A}_l. 
\end{split}
\end{equation}

We can show that the SPE of the game is the steady state of the ODE dynamics in \eqref{eq:systemDynaimcs}, \eqref{eq:attackDynaimcs},  and the interior stationary points of the dynamics are the SPE of the game \cite{zhu2013game}. 

\subsubsection{Heterogeneous and Hybrid Learning}
The entropy regulation terms in \eqref{eq:SP} and \eqref{eq:AP} result in a closed form of strategies and  learning dynamics in \eqref{eq:policyLearning}. 
Without the closed form, distributed learners can adopt general learning schemes which combine the payoff and the strategy update as stated in \cite{zhu2010heterogeneous}.
Specifically, algorithm CRL0 mimics the replicator dynamics and updates the strategy according to the current sample value of the utility. 
On the other hand, algorithm CRL1 updates the strategy according to a soft-max function of the estimated utilities so that the most rewarding policy get reinforced and will be picked with a higher probability. 
The first algorithm is  robust yet inefficient, and the second one is fragile yet efficient. 
Moreover, players are not obliged to adopt the same learning scheme at different time. The heterogeneous learning focuses on different players adopting different learning schemes \cite{zhu2010heterogeneous}, while  hybrid learning means that players can choose different  learning schemes at different times based on their rationalities and preferences \cite{zhu2011distributed}. 
According to stochastic approximation techniques, these learning schemes with random updates can be studied using their deterministic ODE counterparts.

\section{Reinforcement Learning for Uncertain Environments}
\label{sec:honeypot}
\begin{figure*}[h]
\sidecaption[t]
\centering
\includegraphics[width=0.63 \textwidth]{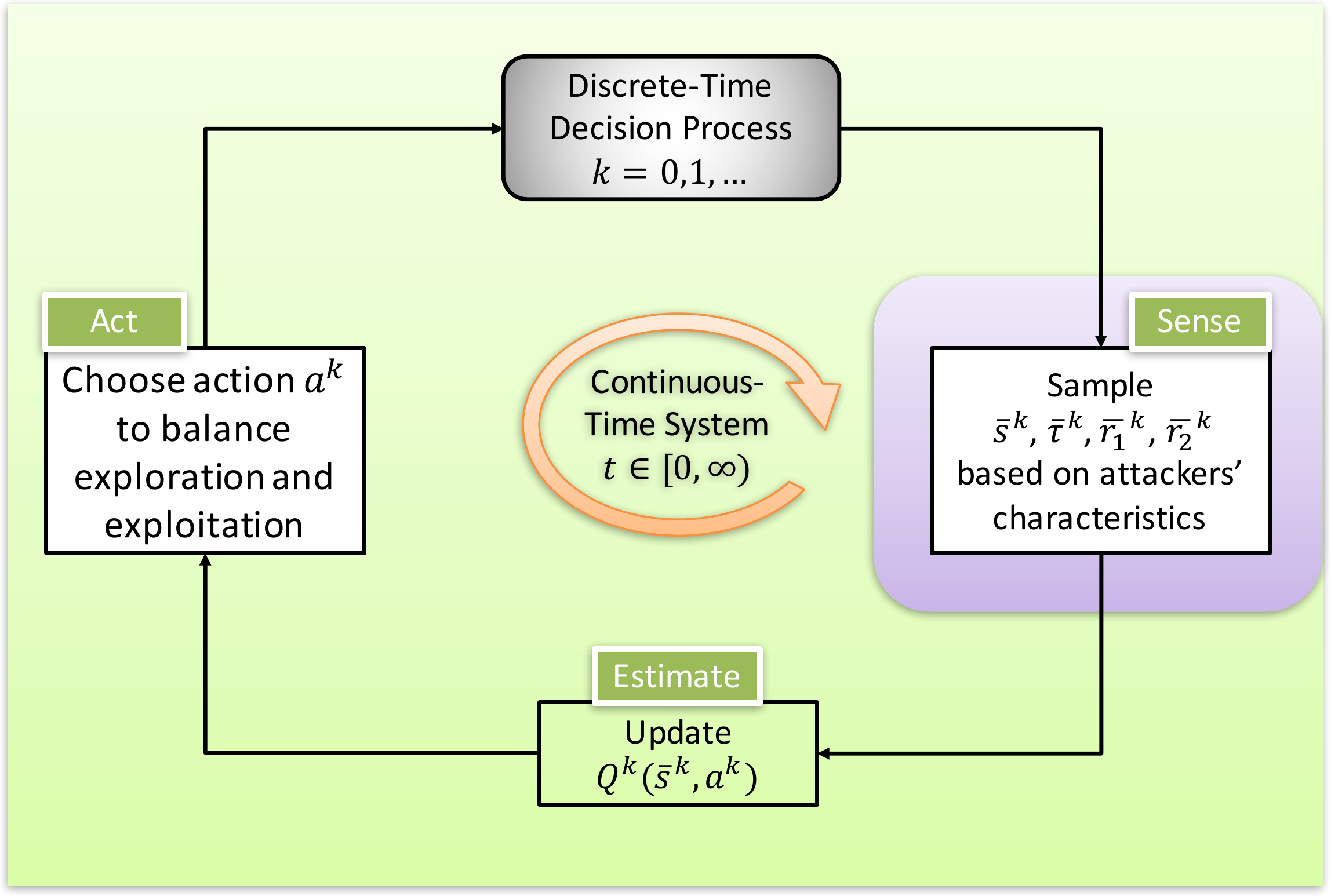}
\caption{
The feedback structure of reinforcement learning methods on SMDP. The red background means that the attacker's characteristics determine the environmental uncertainties and the samples observed in the honeynet. The attacker is not involved in parts of the green background. 
The learning scheme in Fig. \ref{fig: Feedback3} extends  the one in Section \ref{sec:MTD} to consider a continuous time elapse and multistage transitions. 
}
\label{fig: Feedback3}
\end{figure*}

This section considers uncertainties on the entire environment, i.e., the state transition, the sojourn time, and the investigation payoff, in the active defense scenario of the honeypot engagement \cite{huangHoneypot}. 
We use the Semi-Markov Decision Process (SMDP) to capture these environmental uncertainties in the continuous time system. 
Although the attacker's duration time is continuous at each honeypot, the defender's engagement action is applied at a discrete time epoch. 
Based on the observed samples at each decision epoch, the defender can estimate the environment elements determined by attackers' characteristics, and use reinforcement learning methods to obtain the optimal policy. 
We plot the entire feedback learning structure in Fig. \ref{fig: Feedback3}. 
Since the attacker should not identify the existence of the honeypot and the defender's engagement actions, he will not take actions to jeopardize the learning. 

\subsection{Honeypot Network and SMDP Model}
\begin{figure}[h]
\centering
\includegraphics[width=1 \textwidth]{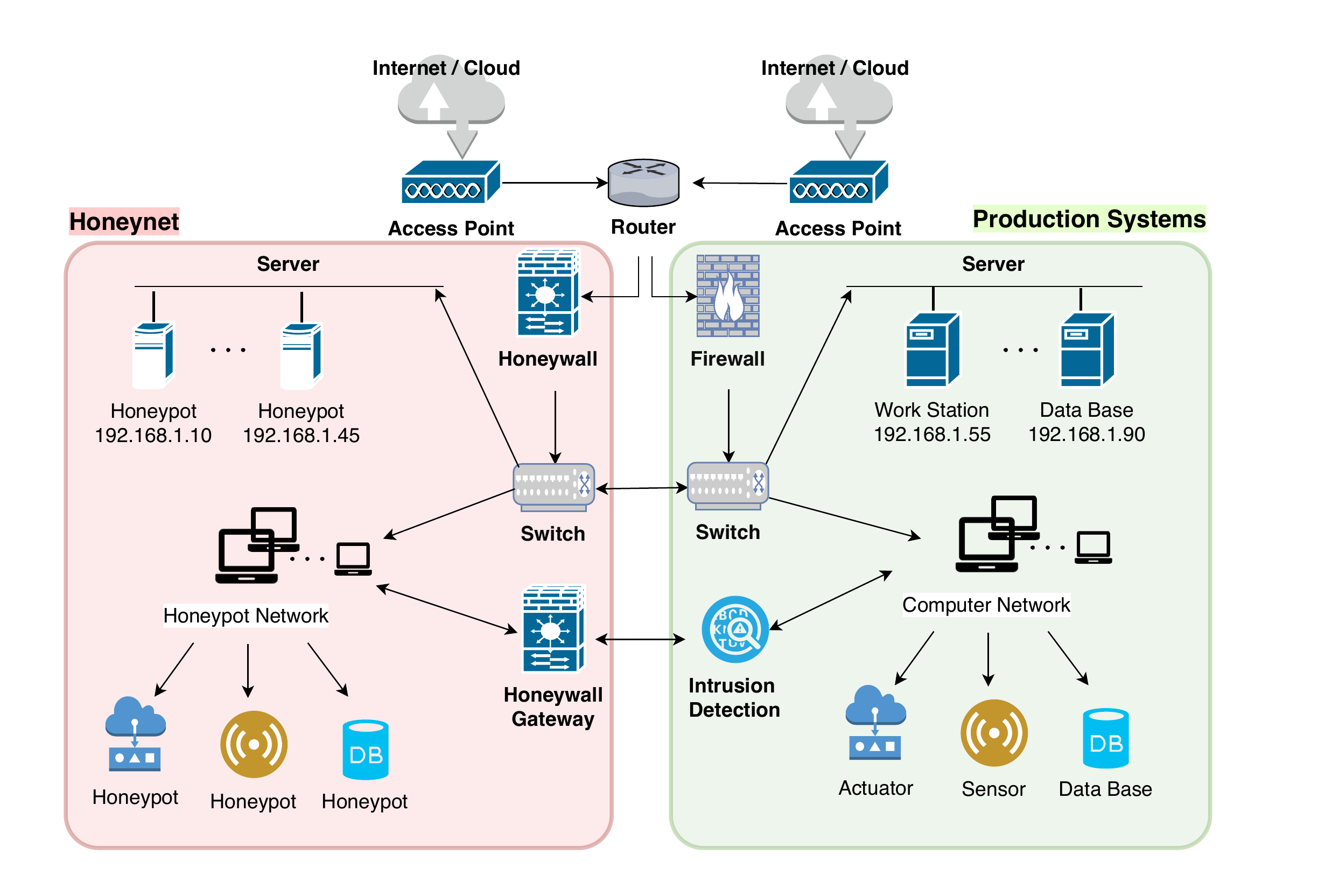}
\caption{
The honeynet in red emulates and shares the same structure as the targeted production system in green. 
 }
 \label{fig: SystemStructure}
\end{figure}

\begin{figure}[]
\sidecaption[t]
%\centering
\includegraphics[width=0.63\textwidth]{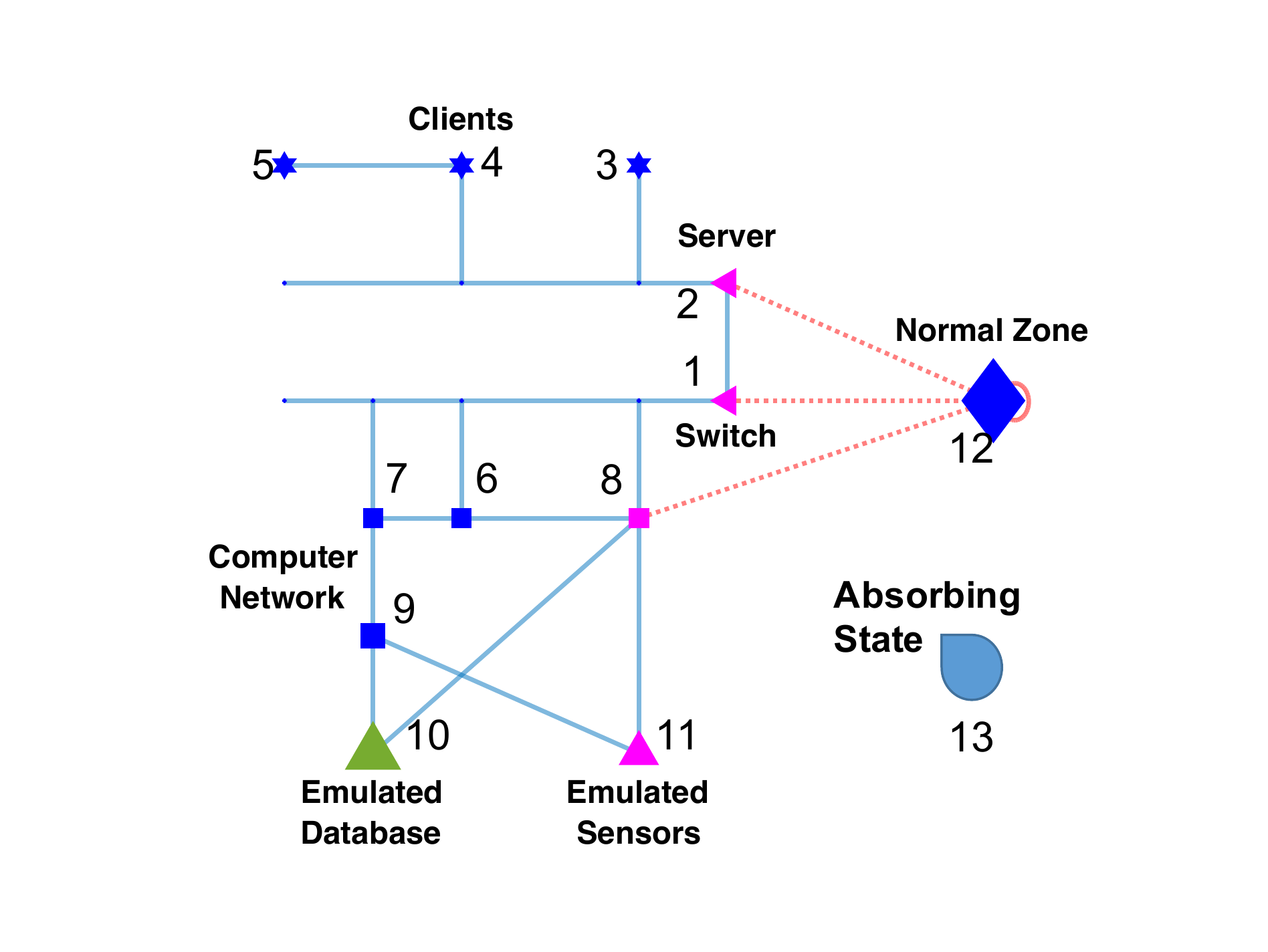}
\caption{
Honeypots emulate different components of the production system.  
Actions $a_E,a_P,a_L,a_H$ are denoted in red, blue, purple, and green, respectively. The size of node $n_i$ represents the state value $v(s_i), i\in \{1,2,\cdots,11\}$.  
}
 \label{fig: SMDPstructure}
\end{figure}

The honeypots form a network to emulate a production system. From an attacker's viewpoint, two network structures are the same as shown in Fig. \ref{fig: SystemStructure}. 
Based on the network topology, we introduce the continuous-time infinite-horizon discounted SMDPs, which can be summarized by the tuple 
$ \{t\in [0,\infty), \mathcal{S},
\mathcal{A}({s_j}), 
tr(s_l| s_j, a_j),  \allowbreak 
z(\cdot| s_j,a_j,s_l), r^{\gamma}( s_j,a_j,s_l), \gamma\in [0,\infty)\}$. 
We illustrate each element of the tuple through a $13$-state example in Fig. \ref{fig: SMDPstructure}.

%\subsubsection{State and State-Dependent Action}
Each node in Fig. \ref{fig: SMDPstructure} represents a state $s_i\in \mathcal{S}, i\in \{1,2,\cdots,13\}$. 
At time $t\in [0,\infty)$, the attacker is either at one of the honeypot node denoted by state $s_i\in \mathcal{S}, i\in \{1,2,\cdots,11\}$, at the normal zone $s_{12}$, or at a virtual absorbing state $s_{13}$ once attackers are ejected or terminate on their own. 
At each state $s_i\in \mathcal{S}$,  the defender can choose an action $a_i\in \mathcal{A}(s_i)$. 
For example, at  honeypot nodes, the defender can conduct action $a_E$ to eject the attacker, action $a_P$ to purely record the attacker's activities, low-interactive action  $a_L$, or high-interactive action $a_H$, i.e., $\mathcal{A}(s_i):=\{a_E, a_P,a_L,a_H\}, i\in \{1,\cdots,
\allowbreak
N\}$. 
The high-interactive action is costly to implement yet can both increases the probability of a longer sojourn time at honeypot $n_i$, and reduces the probability of attackers penetrating the normal system from $n_i$ if connected. 
If the attacker resides in the normal zone either from the beginning or later through the pivot honeypots, the defender can choose either action $a_E$ to eject the attacker immediately, or action $a_A$ to attract the attacker to the honeynet by generating more deceptive inbound and outbound traffics in the honeynet, i.e., $\mathcal{A}(s_{12}):=\{a_E,a_A\}$.  
%For example, a sudden increases of traffic during the attacker's stay may ware him. 
%\subsubsection{Continuous-Time Process and Discrete Decision Model}
%\label{sec:transition}
Based on the current state $s_j\in \mathcal{S}$ and the defender's action $a_j\in \mathcal{A}(s_j)$, the attacker transits to state $s_l\in \mathcal{S}$ with probability $tr(s_l|s_j,a_j)$ and the sojourn time at state $s_j$ is a continuous random variable with probability density $z(\cdot| s_j,a_j, s_l)$. 
Once the attacker arrives at a new honeypot $n_i$, the defender dynamically applies an interaction action at  honeypot $n_i$ from $\mathcal{A}(s_i)$ and keeps interacting with the attacker until she transits to the next honeypot. 
If the defender changes the action before the transition,  the attacker may be able to detect the change and become aware of the honeypot. 
Since the decision is made at the time of transition, we can transform the above continuous time model on horizon $t\in [0,\infty)$ into a discrete decision model at decision epoch $k\in \{0,1,\cdots, \infty\}$. 
The time of the attacker's $k^{th} $ transition is denoted by a random variable $T^k$, the landing state is denoted as $s^k\in \mathcal{S}$, and the adopted action  after arriving at $s^k$ is denoted as $a^k\in \mathcal{A}(s^k)$. 

%\subsubsection{Investigation Reward Maximization}
%\label{sec:investigationreward}
The defender gains an investigation reward by engaging and analyzing the attacker in the honeypot. 
To simplify the notation, we segment the investigation reward during time $t\in [0,\infty)$ into ones at discrete decision epochs $T^k, k\in \{0,1,\cdots, \infty\}$. 
When $\tau\in [T^k,T^{k+1}]$ amount of time elapses at stage $k$, the defender's investigation reward 
$
r(s^k,a^k,s^{k+1},T^k,T^{k+1},\tau)=r_1(s^k,a^k,s^{k+1})\mathbf{1}_{\{\tau=0\}}+r_2(s^k,a^k, T^k,T^{k+1},\tau)
$, at time $\tau$ of stage $k$, is the sum of two parts. The first part is the immediate cost of applying engagement action $a^k \in \mathcal{A}(s^k)$ at state $s^k \in \mathcal{S}$ and the second part is the reward rate of threat information acquisition minus the cost rate of persistently generating deceptive traffics. 
Due to the randomness of the attacker's behavior, the information acquisition can also be random, thus the actual reward rate $r_2$ is perturbed by an additive zero-mean noise $w_r$. 
As the defender spends longer time interacting with attackers, investigating their behaviors and acquires better understandings of their targets and TTPs, less new information can be extracted. 
In addition, the same intelligence becomes less valuable as time elapses due to the  timeliness. 
Thus, we use a discounted factor of $\gamma\in [0,\infty)$ to penalize the decreasing value of the investigation reward as time elapses.

The defender aims at a policy $\pi\in \Pi$ which maps state $s^k\in \mathcal{S}$ to action $a^k\in \mathcal{A}(s^k)$ to maximize the long-term expected utility starting from state $s^0$, i.e., 
\begin{equation}
u(s^0,\pi)=E[\sum_{k=0}^{\infty} \int_{T^k}^{T^{k+1}} e^{-\gamma(\tau+T^k)} (r(S^k,A^k,S^{k+1},T^k,T^{k+1},\tau)+w_r)d\tau]. 
\end{equation}

At each decision epoch, the value function $v(s^0)=\sup_{\pi \in \Pi }u(s^0,\pi)$ can be represented by dynamic programming, i.e., 
\begin{equation}
\label{eq:DPgeneral}
v(s^0)=\sup_{a^0\in \mathcal{A}(s^0)} E[\int_{T^0}^{T^{1}} e^{-\gamma (\tau+T^0)}r(s^0,a^0,S^{1},T^0,T^{1},\tau)d\tau+e^{-\gamma T^1}v(S^1)]. 
\end{equation}

We assume a constant reward rate $r_2(s^k,a^k,T^k,T^{k+1},\tau)=\bar{r}_2(s^k,a^k)$ for simplicity. 
Then, \eqref{eq:DPgeneral} can be transformed into an  equivalent  MDP form, i.e., $ \forall s^0\in \mathcal{S}$, 
\begin{equation}
v(s^0)=\sup_{a^0\in \mathcal{A}(s^0)} \sum_{s^1\in \mathcal{S}} tr(s^1|s^0,a^0) (r^{\gamma}(s^0,a^0,s^1)+z^{\gamma}(s^0,a^0,s^1)v(s^1)), 
\end{equation}
where ${z^{\gamma}}(s^0,a^0,s^1):=\int_0^{\infty}e^{-\gamma \tau} z(\tau|s^0,a^0,s^1)d\tau\in [0,1]$ is the Laplace transform of the sojourn probability density $z(\tau|s^0,a^0,s^1)$ and the equivalent reward 
$r^{\gamma}(s^0,a^0,s^1)\allowbreak
:=r_1(s^0,a^0,s^1)+\frac{\bar{r}_2(s^0,a^0)}{\gamma} (1-z^{\gamma}(s^0,a^0,s^1))\in [-m_c,m_c]$ is assumed to be bounded by a constant $m_c$. 

\begin{definition}
\label{def:regularcondition}
There exists constants $\theta\in (0,1)$ and $\delta>0$ such that 
\begin{equation}
\sum_{s^1\in \mathcal{S}} tr(s^1|s^0,a^0) z(\delta |s^0,a^0,s^1)\leq 1-\theta , \forall s^0\in \mathcal{S}, a^0 \in \mathcal{A}(s^0). 
\end{equation}
\qed
\end{definition}

The right-hand side of \eqref{eq:DPgeneral} is a contraction mapping under the regulation condition in Definition \ref{def:regularcondition}.
Then, we can find the unique optimal policy $\pi^*=arg\max_{\pi \in \Pi }u(s^0,\pi)$ by value iteration, policy iteration or linear programming. 
Fig. \ref{fig: SMDPstructure} illustrates the optimal policy  and the state value  by the color and the size of the node, respectively. 
In the example scenario, the honeypot of database $n_{10}$ and  sensors  $n_{11}$ are the main and secondary targets of the attacker, respectively. Thus, defenders can obtain a higher investigation reward when they manage to engage the attacker in these two honeypot nodes with a larger probability and for a longer time. 
However, instead of naively adopting high interactive actions,  
a savvy defender also balances the high implantation cost of $a_H$. Our quantitative results indicate that the high interactive action should only be applied at $n_{10}$ to be cost-effective. 
On the other hand, although the bridge nodes $n_1,n_2,n_8$ which connect to the normal zone $n_{12}$ do not contain higher investigation rewards than other nodes, the defender still takes action $a_L$  at these nodes. 
The goal is to either increase the probability of attracting attackers away from the normal zone or reduce the probability of attackers penetrating the normal zone from these bridge nodes.

\subsection{Reinforcement Learning of SMDP} 
\label{sec:reinforcementlearning}
The absent knowledge of the attacker's characteristics results in environmental uncertainty of the investigation reward, the attacker's transition probability, and the sojourn distribution. 
We use $Q$-learning algorithm to obtain the optimal  engagement policy based on the actual experience of the honeynet interactions, i.e., $\forall \bar{s}^k\in \mathcal{S}, \forall a^k\in \mathcal{A}(\bar{s}^k)$, 
\begin{equation}
\label{eq:Qlearning}
\begin{split}
Q^{k+1}(\bar{s}^k,a^k)=&(1-\alpha^k(\bar{s}^k,a^k))Q^{k}(\bar{s}^k,a^k)+ \alpha^k(\bar{s}^k,a^k)[ \bar{r}_1(\bar{s}^k,a^k,\bar{s}^{k+1})
\\
&
+\bar{r}_2(\bar{s}^k,a^k)\frac{(1-e^{-\gamma \bar{\tau}^k})}{\gamma}-e^{-\gamma \bar{\tau}^k}\max_{a'\in \mathcal{A}(\bar{s}^{k+1})} Q^k(\bar{s}^{k+1},a')], 
\end{split}
\end{equation}
where $\alpha^k(\bar{s}^k,a^k)\in (0,1)$ is the learning rate, $\bar{s}^k, \bar{s}^{k+1}$ are the observed states at stage $k$ and $k+1$, $\bar{r}_1,\bar{r}_2$ is the observed investigation rewards, and $\bar{\tau}^k$ is the observed sojourn time at state $s^k$. 
When the learning rate satisfies the condition of convergency in Definition \ref{def:convergencyCondition}, i.e., 
$\sum_{k=0}^\infty \alpha^k(s^k,a^k)=\infty, \sum_{k=0}^\infty (\alpha^k(s^k,a^k))^2<\infty, \forall s^k\in \mathcal{S}, \forall a^k\in \mathcal{A}(s^k)$, and all state-action pairs are explored infinitely, $\max_{a'\in \mathcal{A}(s^k)} \allowbreak
Q^{\infty}(s^{\infty},a')$,  in \eqref{eq:Qlearning} converges to value $v(s^k)$ with probability $1$. 

At each decision epoch $k\in \{0,1,\cdots\}$, the action $a^k$ is chosen according to the $\epsilon$-greedy policy, i.e., the defender chooses the optimal action $arg\max_{a'\in \mathcal{A}(s^k)} Q^k(s^k,a')$ with a probability $1-\epsilon$, and a random action with a probability $\epsilon$. 
Note that the exploration rate $\epsilon\in (0,1]$ should not be too small to guarantee sufficient samples of all state-action pairs. The $Q$-learning algorithm under a pure exploration policy $\epsilon=1$ still converges yet at a slower rate. 

\begin{figure}
\sidecaption[t]
%\centering
\includegraphics[width=0.62 \textwidth]{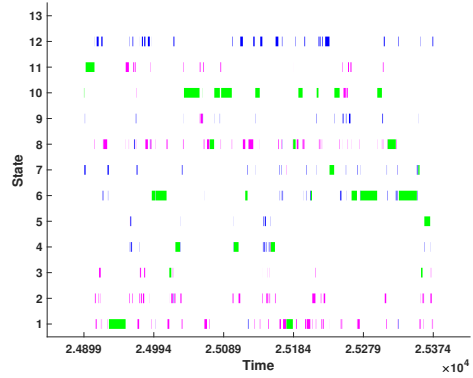}
\caption{
One instance of $Q$-learning on SMDP where the $x$-axis shows the sojourn time and the $y$-axis represents the state transition. 
The chosen actions $a_E,a_P,a_L,a_H$ are denoted in red, blue, purple, and green, respectively. 
}
 \label{fig: SamplePath}

\end{figure}

In our scenario, the defender knows the reward of ejection action $a_A$ and $v(s_{13})=0$, thus does not need to explore action $a_A$ to learn it. 
We plot one learning trajectory of the state transition and sojourn time under the  $\epsilon$-greedy exploration policy in Fig. \ref{fig: SamplePath}, where the chosen actions $a_E,a_P,a_L,a_H$ are denoted in red, blue, purple, and green, respectively. 
If the ejection reward is unknown, the defender should be restrictive in exploring $a_A$ which terminates the learning process. Otherwise, the defender may need to  engage with a group of attackers who share similar behaviors to obtain sufficient samples to learn the optimal engagement policy. 

\begin{figure}[h]
    \centering
    \begin{subfigure}[]{0.45\textwidth}
        \centering
        \includegraphics[width=1 \textwidth]{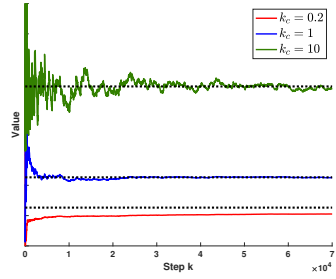}
        \caption{The convergence rate under different values of $k_c$.  }
        \label{fig: compare_kc}
    \end{subfigure}%
    \hfill 
    \begin{subfigure}[]{0.45\textwidth}
        \centering
        \includegraphics[width=1 \textwidth]{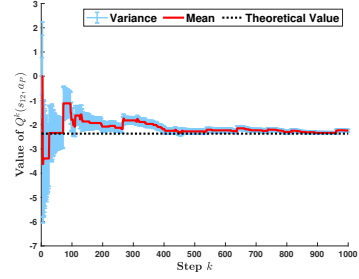}
        \caption{The evolution of the mean and the variance of $Q^k(s_{12},a_P)$. }
         \label{fig:Sample100}
    \end{subfigure}
    \caption{
    Convergence results of $Q$-learning over SMDP. 
 }
  \label{fig: Sample100}
\end{figure}

In particular, we choose $\alpha^k(s^k,a^k)=\frac{k_c}{k_{\{s^k,a^k\}}-1+k_c}, \forall s^k\in \mathcal{S}, \forall a^k\in \mathcal{A}(s^k)$, to guarantee the asymptotic convergence, where $k_c \in (0,\infty)$ is a constant parameter and  $k_{\{s^k,a^k\}}\in \{0,1,\cdots\}$ is the number of visits to state-action pair $\{s^k,a^k\}$ up to stage $k$. 
We need to choose a proper value of $k_c$ to guarantee a good numerical performance of convergence in finite steps as shown in Fig. \ref{fig: compare_kc}. We shift the green and blue lines vertically to avoid the overlap with the red line and 
represent the corresponding theoretical values in dotted black lines. 
If $k_c$ is too small as shown in the red line, the learning rate decreases so fast  that new observed samples hardly update the $Q$-value and the defender may need a long time to learn the right value. 
However, if $k_c$ is too large as shown in the green line, the learning rate decreases so slow that new samples contribute significantly to the current $Q$-value. It causes a large variation and a slower convergence rate of $\max_{a'\in \mathcal{A}(s_{12})} Q^k(s_{12},a')$. 

We show the convergence of the policy and value under $k_c=1,\epsilon=0.2$, in the video demo (See URL: \url{https://bit.ly/2QUz3Ok}). 
In the video, the color of each node $n^k$ distinguishes the defender's action $a^k$ at state $s^k$ and the size of the node is proportional to $\max_{a'\in \mathcal{A}(s^k)} Q^k(s^k,a')$ at stage $k$. 
To show the convergence, we decrease the value of $\epsilon$ gradually to $0$ after $5000$ steps. 
Since the convergence trajectory is stochastic, we run the simulation for $100$ times and plot the mean and the variance of $Q^k(s_{12}, a_P)$ of state $s_{12}$ under the optimal policy $\pi(s_{12})=a_P$ in Fig. \ref{fig: Sample100}. The mean in red converges to the theoretical value in about $400$ steps and the variance in blue reduces dramatically as step $k$ increases. 

\section{Conclusion and Discussion} 
\label{sec:conclusion}
 This chapter has introduced three defense schemes, i.e., defensive deception to detect and counter adversarial deception, feedback-driven Moving Target Defense (MTD) to increase the attacker's probing and reconnaissance costs, and adaptive honeypot engagement to gather fundamental threat information. 
 These schemes satisfy the Principle of 3A Defense as they actively protect the system prior to the attack damages, provide strategic defenses autonomously, and apply learning to adapt to uncertainty and changes.  
These schemes possess three progressive levels of information restrictions, which lead to different strategic learning schemes to estimate the parameter, the payoff, and the environment. 
All these learning schemes, however, have a feedback loop to sense samples, estimate the unknowns, and take actions according to the estimate. 
Our work lays a solid foundation for strategic learning in active, adaptive, autonomous defenses under incomplete information and leads to the following challenges and future directions. 

First, multi-agent learning in non-cooperative environments is challenging due to the coupling and interaction between these heterogeneous agents. 
The learning results depend on all involving agents yet other players' behaviors, levels of rationality, and learning schemes are not controllable and may change abruptly.  
Moreover, as attackers become aware of the active defense techniques and the learning scheme under incomplete information, the savvy attacker can attempt to interrupt the learning process. 
For example, attackers may sacrifice their immediate rewards and take incomprehensible actions instead so that the defender learns incorrect attack characteristics. 
The above challenges motivate robust learning methods under non-cooperative and even adversarial environments. 

Second, since the learning process is based on samples from real interactions, the defender needs to concern the system safety and security during the learning period, while in the same time, attempts to achieve more accurate learning results of the attack's characteristics.  
Moreover, since the learning under non-cooperative and  adversarial environments may terminate unpredictably at any time, the asymptotic convergence would not be critical for security. 
The defender needs to care more about the time efficiency of the learning, i.e., how to achieve a sufficiently good estimate in a finite number of steps. 

Third, instead of learning from scratch, the defender can attempt to reuse the past experience with attackers of similar behaviors to expedite the learning process, which motivates the investigation of transfer learning in reinforcement learning \cite{taylor2009transfer}. Some side-channel information may also contribute to the learning to allow agents to learn faster.

\bibliographystyle{IEEEtran}
\bibliography{BayesianRiccati}

\end{document}